\pdfoutput=1
\documentclass[lettersize,journal]{IEEEtran}
\usepackage{amsmath,amsfonts}
\usepackage{algorithmic}
\usepackage{algorithm}
\usepackage{array}
\usepackage{textcomp}
\usepackage{stfloats}
\usepackage{url}
\usepackage{verbatim}
\usepackage{graphicx}
\usepackage{cite}
\usepackage{acronym}
\usepackage{caption}
\usepackage{subcaption}
\usepackage[table,xcdraw]{xcolor}
\usepackage{tikz}
\newcommand*\circled[1]{\tikz[baseline=(char.base)]{
            \node[shape=circle,draw,inner sep=1pt] (char) {#1};}}

\begin{document}

\newcommand{\new}[1]{\noindent{\color{blue}\textbf{}~#1}}
\newcommand{\revise}[1]{\noindent{\color{orange}\textbf{}~#1}}
\newcommand{\todo}[1]{\noindent\textit{\color{red}\textbf{}~#1}}

\acrodef{ML}{Machine Learning}
\acrodef{DL}{Deep Learning}
\acrodef{IoT}{Internet-of-Things}
\acrodef{DNN}{Deep Neural Network}
\acrodef{FP}{Floating Point}
\acrodef{SoA}{State-of-the-Art}
\acrodef{ISA}{Instruction Set Architecture}
\acrodef{PULP}{Parallel Ultra-Low-Power}
\acrodef{SoC}{System on Chip}
\acrodef{DWE}{Depth-Wise Convolution Engine}
\acrodef{TPE}{Tensor Product Engine}
\acrodef{HWPE}{Hardware Processing Engine}
\acrodef{TCDM}{Tightly-Coupled Data Memory}
\acrodef{HCI}{Heterogeneous Cluster Interconnect}
\acrodef{NN-RF}{Neural Network Register File}
\acrodef{GP-RF}{General-Purpose Register File}
\acrodef{FMA}{Fused Multiply-Add}
\acrodef{HWC}{Height-Width-Channel}
\acrodef{AE}{Autoencoder}


\title{\textsc{Darkside}: A Heterogeneous RISC-V Compute Cluster for Extreme-Edge On-Chip DNN Inference and Training}

\author{\IEEEauthorblockN{
Angelo Garofalo\IEEEauthorrefmark{1},
Yvan Tortorella\IEEEauthorrefmark{1},
Matteo Perotti\IEEEauthorrefmark{2},
Luca Valente\IEEEauthorrefmark{1}, \\
Alessandro Nadalini\IEEEauthorrefmark{1}, 
Luca Benini\IEEEauthorrefmark{1}\IEEEauthorrefmark{2},
Davide Rossi\IEEEauthorrefmark{1},
 and
Francesco Conti\IEEEauthorrefmark{1}}
\thanks{This  work  was  supported  in  part  by  EU  Horizon  2020  Research  and  In-novation  projects  The  European  Pilot  under  Grant  101034126,  in  part  by WiPLASH  under  Grant  863337,  in  part  by  ECSEL  Horizon  2020  project AI4DI under Grant 826060 and in part by GreenWaves Technologies.}
 
\IEEEauthorblockA{\IEEEauthorrefmark{1}Department of Electrical, Electronic and Information Engineering (DEI), University of Bologna, Italy}

\IEEEauthorblockA{\IEEEauthorrefmark{2}IIS Integrated Systems Laboratory, ETH Zurich, Switzerland}}   

\markboth{Journal of \LaTeX\ Class Files,~Vol.~14, No.~8, August~2021}%
{Shell \MakeLowercase{\textit{et al.}}: A Sample Article Using IEEEtran.cls for IEEE Journals}

\maketitle
\begin{abstract}
On-chip DNN inference and training at the Extreme-Edge (\textit{TinyML}) impose strict latency, throughput, accuracy and flexibility requirements. Heterogeneous clusters are promising solutions to meet the challenge, combining the flexibility of DSP-enhanced cores with the performance and energy boost of dedicated accelerators.
%
%
We present \textsc{Darkside}, a System-on-Chip with a heterogeneous cluster of 8~RISC-V cores enhanced with 2-b to 32-b~mixed-precision integer arithmetic. To boost performance and efficiency on key compute-intensive Deep Neural Network~(DNN) kernels, the cluster is enriched with three digital accelerators: a specialized engine for low-data-reuse depthwise convolution kernels (up to 30~MAC/cycle); a minimal overhead datamover to marshal 1-b to 32-b~data on-the-fly; a 16-b floating point Tensor Product Engine~(TPE) for tiled matrix-multiplication acceleration. \textsc{Darkside} is implemented in 65nm CMOS technology. The cluster achieves a peak integer performance of 65~GOPS and a peak efficiency of 835~GOPS/W when working on 2-b~integer DNN kernels. When targeting floating-point tensor operations, the TPE provides up to 18.2~GFLOPS of performance or 300~GFLOPS/W of efficiency -- enough to enable on-chip floating-point training at competitive speed coupled with ultra-low power quantized inference.

\end{abstract}


\section{Introduction}
The recent mega-trend aiming at deploying \ac{ML} and \ac{DL} at the extreme edge of the \ac{IoT}, usually referred to as Tiny Machine Learning (\textit{TinyML}), has reached outstanding results. For example, MobileNets~\cite{sandler2018mobilenetv2} have rapidly become state-of-the-art compute workloads used for classification and object detection inference tasks, but also as a flexible template for tasks not related to vision~\cite{cai2018proxylessnas,incze2018bird, zhang2019lightweight}. 

Next-generation TinyML \ac{IoT} devices, however, will likely require also the capability to adapt the deployed \ac{DL} model to new data directly in the field. Re-training the model on data centers with data collected on-field from the distributed \ac{IoT} end-nodes might be expensive in terms of latency and power and inconvenient from the privacy and security viewpoints. Therefore, a common direction of TinyML is to rethink the deployed \ac{DNN} as a dynamic model that can adapt by learning from newly sensed data directly on the device. Recent progress in this research area concerns \ac{DNN} model tuning, partial on-chip training~\cite{ravaglia2021tinyml} or unsupervised continual learning~\cite{ren2021tinyol}, which have been applied successfully to many \ac{IoT} applications, such as anomaly detection tasks~\cite{unknown}. 


Satisfying both the needs of TinyML inference and on-device adaptation requires devices that are highly flexible and efficient simultaneously on these two very different tasks.
Inference in TinyML devices typically adopts low-bitwidth integer arithmetic, relying on well-established Quantization-aware training~\cite{jacob2018quantization} and post-training quantization techniques~\cite{hubara2017quantized}.
Mixed-precision approaches~\cite{rusci2020leveraging,schaefer2022edge}, where the activations and the weights of all \ac{DNN} layers can be quantized with different precisions, are \ac{SoA} solutions to reduce the accuracy drop compared to full-precision models (e.g., within a 3 to 6\% range in ImageNet Top-1), while cutting the model footprint by a significant factor ($\sim$7$\times$ on MobileNets~\cite{rusci2020leveraging}). 

Specialized digital accelerators like~\cite{chen2019eyeriss, moons201714, lee2018unpu, desoli201714} achieve outstanding performance (1-50 TOPS/W) and energy efficiency (10-100 TOPS/W) on \ac{DNN} kernels by exploiting low-bitwidth integer arithmetic. Recently this approach has also been adopted in analog-digital mixed-signal solutions~\cite{khaddam2022hermes, ueyoshi2022diana}, boosting energy efficiency up to hundreds of TOPS/W. However, these hardware units are highly specialized in terms of supported functionality and numerical precision and leak the flexibility needed to adapt to rapidly evolving TinyML models.

A different solution, exploiting clusters of parallel fully programmable architectures, would ensure the highest flexibility while still achieving competitive efficiency by leveraging instruction extensions supporting multiple formats to cover multiple data precision combinations in arithmetic instructions.
Garofalo~et~al.~\cite{garofalo2021xpulpnn} propose parallel RISC-V cores with SIMD sum-of-dot-product instructions and custom mac-load operations to achieve ASIC-like efficiency on symmetric \ac{DNN} convolutions.
To reduce the overhead of instruction decoding for multiple-precision combinations, Ottavi~et~al.~\cite{ottavi2020mixed}
proposed lightweight status-based mixed-precision computing support to a RISC-V processor, showing two orders of magnitude better efficiency than existing commercial microcontroller solutions.

Supporting multiple, mixed-precision computation is not the only flexibility challenge. Unlike the previous generation of TinyML \ac{DNN} models, \ac{SoA} MobileNets and derived networks feature more heterogeneous workloads, with standard convolutions combined with point-wise and depth-wise kernels. Although they have less computation complexity and smaller memory footprint, depthwise layers are characterized by low intrinsic data reuse~\cite{garofalo2022heterogeneous}. For this reason, they are harder to accelerate with massive arrays of processing elements. As a result, in the DNN processing pipeline, Amdahl's effect moves the acceleration bottleneck toward depth-wise kernels. Likewise, data marshalling operations (e.g., low-bitwidth transpose) commonly used in \acp{DNN} heavily rely on sub-byte swap operations, which also contribute to reducing utilization of the arithmetic units.

Introducing on-device training to the picture imposes yet different constraints in terms of performance and footprint, as training has stricter requirements on the data representation: integer arithmetic can not be used due to its limited dynamic range. To develop extreme-edge novel learning algorithms, a decisive effort is underway to adapt learning algorithms to lower-precision like \ac{FP}16 and \ac{FP}8~\cite{rodriguez2018lower, sun2019hybrid}. Despite this, \textit{TinyML} on-chip training workload is still 10-100$\times$ larger than inference~\cite{ravaglia2021tinyml}, and the performance requirements remain very high. Accelerating these workloads with general-purpose processors would require massive cores, blowing up the SoC's area and power consumption unacceptably. Hence, fixed-function custom designs still are the most suitable solutions to deliver significantly high performance within a TinyML compatible area and power budget.

We argue that a single \textit{catch-it-all} solution is infeasible with all these competing requirements. Instead, boosting end-to-end AI-enhanced applications will require \textit{heterogeneous systems} combining different acceleration engines for different kernels, coping with strict power and cost constraints~\cite{VEGA}: multiple programmable cores provide flexible and efficient execution for generic parallel kernels, while specialized hardware accelerators provide extra performance and efficiency boost on essential kernels that dominate the computational workload.

In this work, we present \textsc{Darkside}, a \ac{PULP}-based~\cite{starwars, VEGA} heterogeneous computing \ac{SoC} that targets emerging TinyML inference and on-chip training applications.
We introduce four main innovations in \textsc{Darkside}: \textit{1)} RISC-V cores with advanced low-bitwidth mixed-precision integer computing capabilities; \textit{2)} a \ac{DWE}, \textit{3)} a low-overhead DataMover for marshalling operations, and \textit{4)} a low-power \ac{TPE} for efficient \ac{FP}16 matrix multiplications.
The cores and accelerators are tightly integrated into a shared-L1 cluster to enable advanced hardware/software cooperation. The chip has been fabricated in TSMC 65nm technology and achieves peak integer performance (2-bit) of 65 GOPS at 1.2V with an efficiency of 835 GOPS/W at 0.75V. On \ac{TPE}-accelerated \ac{FP}16 workloads, it achieves up to 18.2 GFLOPS at 1.2V and a peak efficiency of 300 GFLOPS/W and 2.6 GFLOPS at 0.75V, achieving peak performance and efficiency similar to 8-bit integer operations.


\section{\textsc{Darkside} SoC Architecture}

\begin{figure}[t]
    \centering
    \includegraphics[width=0.95\linewidth]{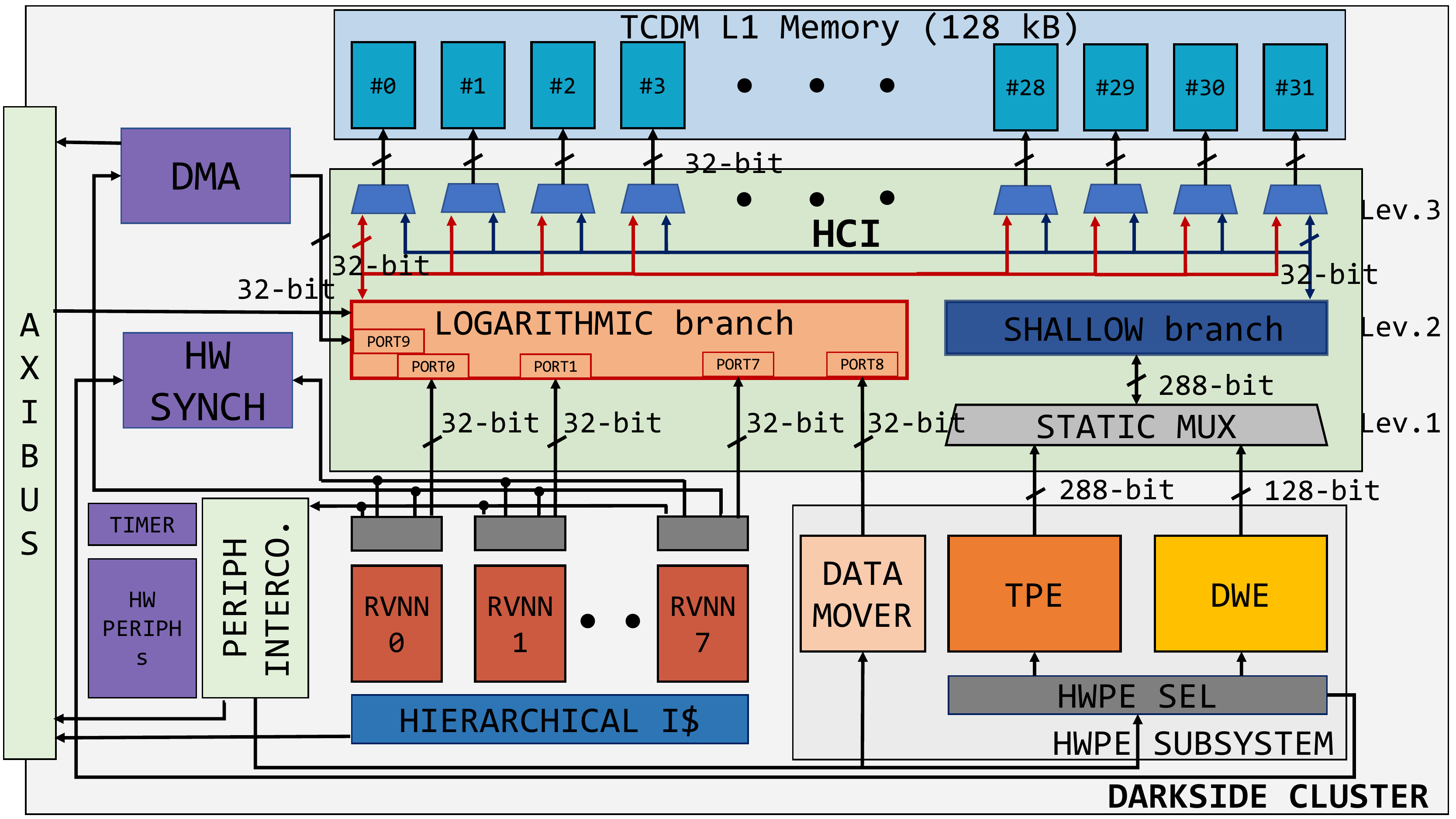}
    \caption{ Overview of the \textsc{Darkside} cluster architecture, featuring 8 RVNN cores, the \ac{TPE}, the \ac{DWE} and the DataMover accelerators.}
    \label{fig:dark_archi}
\end{figure}

\begin{figure}[t]
    \centering
    \includegraphics[width=0.95\linewidth]{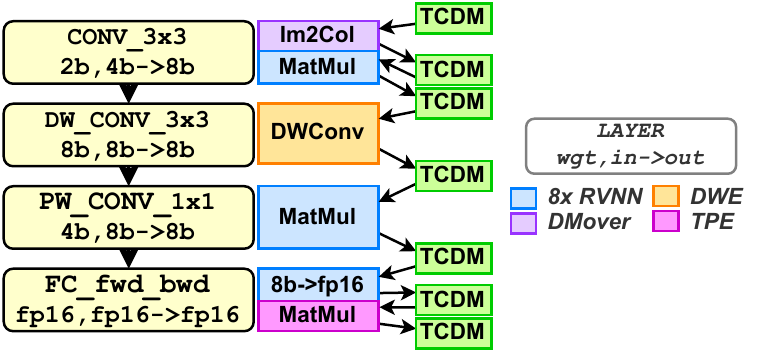}
    \caption{\textsc{Darkside} heterogeneous operation. Figure shows a \ac{DNN} with 3 mixed-precision quantized layers and a final fully-connected layer using all the cluster blocks and communicating through software-managed buffers allocated on the shared L1 memory.}
    \label{fig:heterogeneous_example}
\end{figure}

Fig.~\ref{fig:dark_archi} shows the architecture of the \textsc{Darkside} cluster. It is built around eight 81kGE RISC-V-based 32-bit processors (\textit{RVNN} cores), described in detail in Sec.~\ref{sec:RV-NN}, and three specialized digital accelerators, the \textit{\ac{TPE}}, the \textit{\ac{DWE}} and the \textit{DataMover}. The heterogeneous cluster can be used to support complex ML models, such as those depicted in Fig.~\ref{fig:heterogeneous_example}, through cooperation among its hardware compute units.

To achieve high computing efficiency on a wide range of workloads, the key goal is to minimize the area and power impact of the specialized accelerators integrated into 
\textsc{Darkside}'s cluster and improve their efficiency on data movements.

To save area, we design the three accelerators to have small internal buffers, the minimal necessary to guarantee datapath utilization rate close to 100\%, while they use the 128 kB scratch-pad multi-banked L1 \ac{TCDM} of the cluster as primary data buffer. Moreover, to minimize their power consumption, especially when the accelerators are not used, we added clock gating cells and operand isolation gates. This strategy cut the dynamic power consumption of the \textit{idle} accelerators with minimal additional logic.
To improve the performance and the energy efficiency of the accelerators in data movements operations, and to east their integration into the cluster, each of the three accelerators is incorporated as a Hardware Processing Engine \ac{HWPE}, using a standardized interface~\footnote{https://hwpe-doc.readthedocs.io}. Such an interface exposes to the rest of the cluster a wide data transfer port (typically much wider than 32-bit) to optimize the access to the primary data buffer and a control port that allows the RISC-V cores to program the accelerator through memory-mapped control registers, as visible in Fig.~\ref{fig:dark_archi}. In each \ac{HWPE}, specialized internal \textit{Streamers} move data between the accelerators and the L1 TCDM memory through the data port, converting the memory accesses into data streams to feed the accelerator's datapath. 


The \ac{TCDM} is divided into 32 4-kB SRAM banks, capable of serving 32 requests in parallel and it is shared among the three specialized accelerators and the 8 RVNN general-purpose cores. The memory requests of all the cluster's compute units are routed through a one-cycle latency hierarchical \ac{HCI} (see Section~\ref{sec:hci}), which leverages a request/grant protocol and a world-level interleaving scheme to evenly distribute the requests, minimizing the access contentions toward the SRAM banks. 

The cluster also features a two-level hierarchical instruction cache (I\$), implemented with latch-based SCM to improve the energy efficiency over energy-expensive SRAM cuts. It includes 8 512-B private per-core plus 4kB of two-cycle shared cache to maximize the efficiency with the data-parallel code. A dedicated DMA controller, featuring a similar size as the cores ($\sim$84 kGE), efficiently manages the data transfer between the L2 (off-the-cluster) and L1 memory. The DMA supports 2-D data transfers and up to 16 outstanding transactions, hiding the latency of L2-L1 data transfers on data-intensive kernels~\cite{burrello2021dory}, while saving energy compared to cached-based systems. The cluster integrates also a small Hardware Synchronization Unit ($\sim$30 kGE) which manages fine-grained parallel thread dispatching and clock-gating of idle cores waiting for synchronization, enabling low-overhead and fine-grained parallelism, thus high energy efficiency. 
The cluster resides in a dedicated power and clock domain. It is surrounded by other IPs integrated into a different power and clock domain, namely the \textit{Fabric} domain. The latter includes a controlling RISC-V processor, FLLs for clock generation, a standard set of I/O peripherals and 256 kB of L2 memory containing the code executed by both the compute cluster and the controlling RISC-V core. In the context of this work, the \textit{Fabric} domain serves as a programmable testbench for the cluster, which is the main architectural contribution of this work. The communication bus between the \textit{Fabric} and the cluster domain is AXI4 based, and dual clock first-in-first-out (FIFO) buffers are used for clock domain crossing.

\subsection{Heterogeneous Cluster Interconnect (HCI)}
\label{sec:hci}

To reduce area and simplify the arbitration scheme, the \ac{HCI} is organized hierarchically in three different levels. At the first level, the \ac{TPE} and the \ac{DWE} are \textit{statically multiplexed} to share the same physical \ac{HWPE} 288-bit data port, which is sized to meet the bandwidth requirements of the two accelerators. Since in our computing model, reported in Fig.~\ref{fig:heterogeneous_example}, the \ac{DWE} and the \ac{TPE} are never used concurrently, the static multiplexing strategy is not a concern from a performance perspective. On the contrary, it allows exposing the accelerators' data interface toward the higher levels of the \ac{HCI} with limited area and power costs.

\begin{figure}[t]
    \centering
    \includegraphics[width=\linewidth]{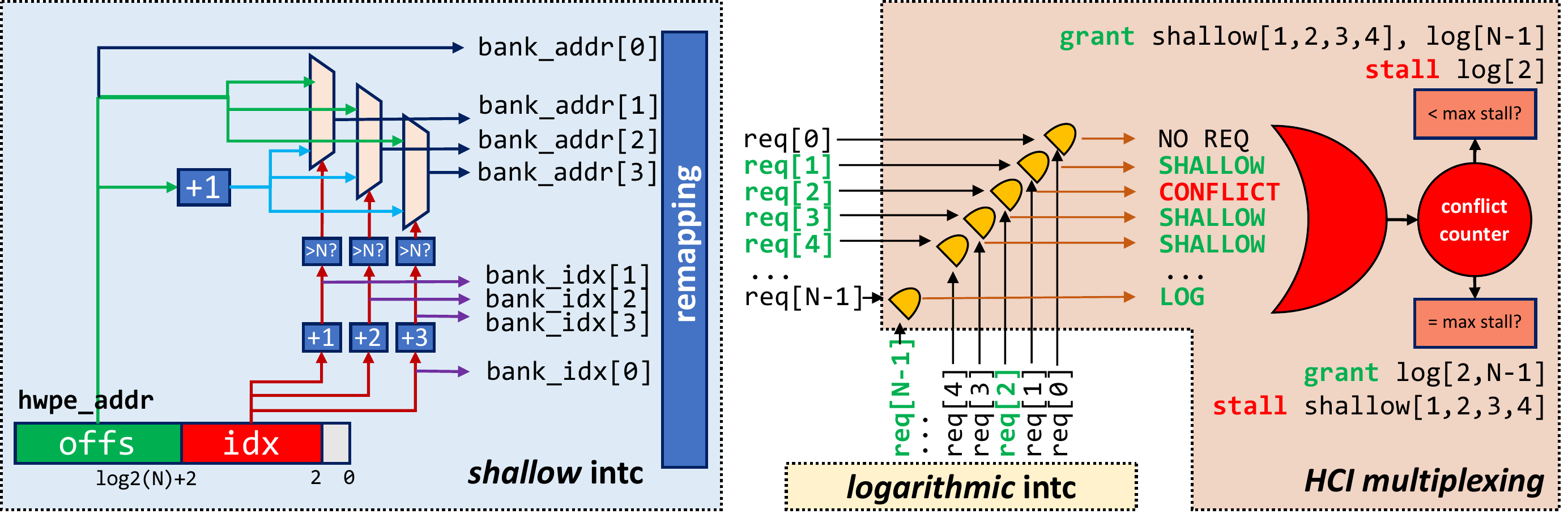}
    \caption{Simplified example of HCI \textit{shallow} routing and arbitration between \textit{shallow} and \textit{logarithmic} branches, considering $N$ TCDM banks. The example shows a \textit{shallow} 128-bit (4-ports) wide access starting on bank 1 and two 32-bit accesses from the \textit{logarithmic} side.}
    \label{fig:hci_rebuttal}
\end{figure}

The second level of the \ac{HCI} is organized in two branches, \textit{logarithmic} and \textit{shallow}, as shown in Fig.~\ref{fig:dark_archi}: the cores, the cluster DMA and the DataMover access the L1 banks from the  \textit{logarithmic} branch through 9 32-bit initiator ports. This branch allows all-to-all single-cycle access from the initiator ports to each word interleaved memory bank. Conflicts are handled by granting one initiator per bank at a time through a round-robin scheme. Instead, the 288-bit muxed data port is connected to the dedicated \textit{shallow} branch, routed to 9 adjacent memory banks without arbitration. Considering a total of $N$ TCDM banks, routing works by splitting the address of the 288-bit wide word in an \textit{index} (bits $2$ through $\mathrm{log}_2(N)+2$) and an \textit{offset} part (upper bits). The index is used to select which TCDM banks are targeted, while the offset is used to compute the bank level address, considering the possibility that the wide word ``rolls over'' the set of banks (if the index corresponds to one of the last banks).

The third level of the \ac{HCI} is at the memory side, where the \ac{TCDM} banks are connected to the two \ac{HCI} branches via multiplexers, granting access to one branch or the other according to a configurable-latency starvation-free rotation scheme. Ports from the logarithmic branch are stalled individually, whereas those from the shallow branch are stalled collectively (a single collision will result in no grant for the whole branch) to reflect the fact that they are actually a single access. Priority is given to a branch configurable via a memory-mapped register, and switched for one access after a configurable number of cycles. Fig.~\ref{fig:hci_rebuttal} showcases this mechanism in an example.

The heterogeneous organization of the interconnect serves two purposes. On the one hand, the \ac{HCI} can be configured in software (by writing a memory-mapped register) to prioritize either the \textit{shallow} or the \textit{logarithmic} branch and guarantee a minimum quality of service (in terms of consecutive stall cycles) to the non-priority branch (by setting a register with the maximum number of stalls that the less priority branch can tolerate). This enables to control and tune the interconnect's performance at a fine granularity. For example, setting priority to the \textit{shallow} branch and maximum stall of 10 cycles in the \textit{logarithmic} branch means that after 10 collisions the priority will be switched to \textit{logaritmic} side for one cycle, hence guaranteeing a 9.1\% collision rate, delivering up to 20.9 GB/s at 290 MHz in the configuration used in \textsc{Darkside} (i.e. 288-bit wide shallow branch and 9 32-bit initiator ports in the logarithmic branch), even on data-intensive kernels.

On the other hand, the scalability of the \textit{logarithmic} interconnect is limited:  attaching the accelerators to a non-hierarchical interconnect would result in a much more complex, larger, and power-hungry interconnect circuit, leading to poor cluster-level performance-per-area and per power. 
The \ac{HCI} occupies 7.3\% ($\sim$220kGE) of the total cluster area; our synthesis trials have shown that the overall the complexity of the interconnect is reduced by 15\% with respect to a purely logarithmic interconnect, which combined with easier timing closure and extended functionality led to the choice of this design.

\subsection{Dynamic Bit-Scalable Fused Mac-Load SIMD Operations}
\label{sec:RV-NN}


\begin{figure*}
     \centering
     \begin{subfigure}[b]{0.45\linewidth}
         \centering
         \includegraphics[width=0.85\linewidth]{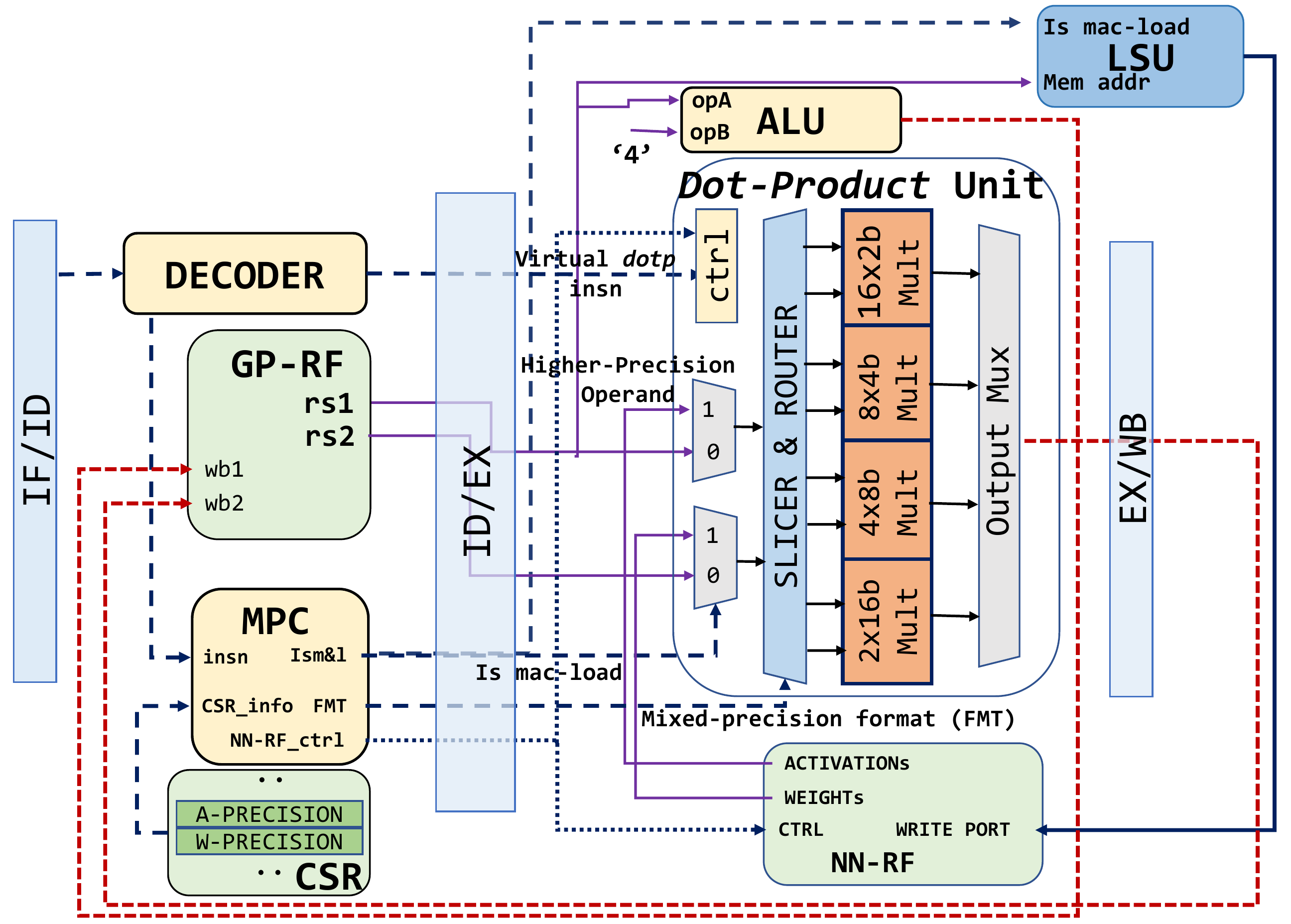}
         \label{fig:y equals x}
     \end{subfigure}
     \quad \quad
     \begin{subfigure}[b]{0.45\linewidth}
         \centering
         \includegraphics[width=0.85\linewidth]{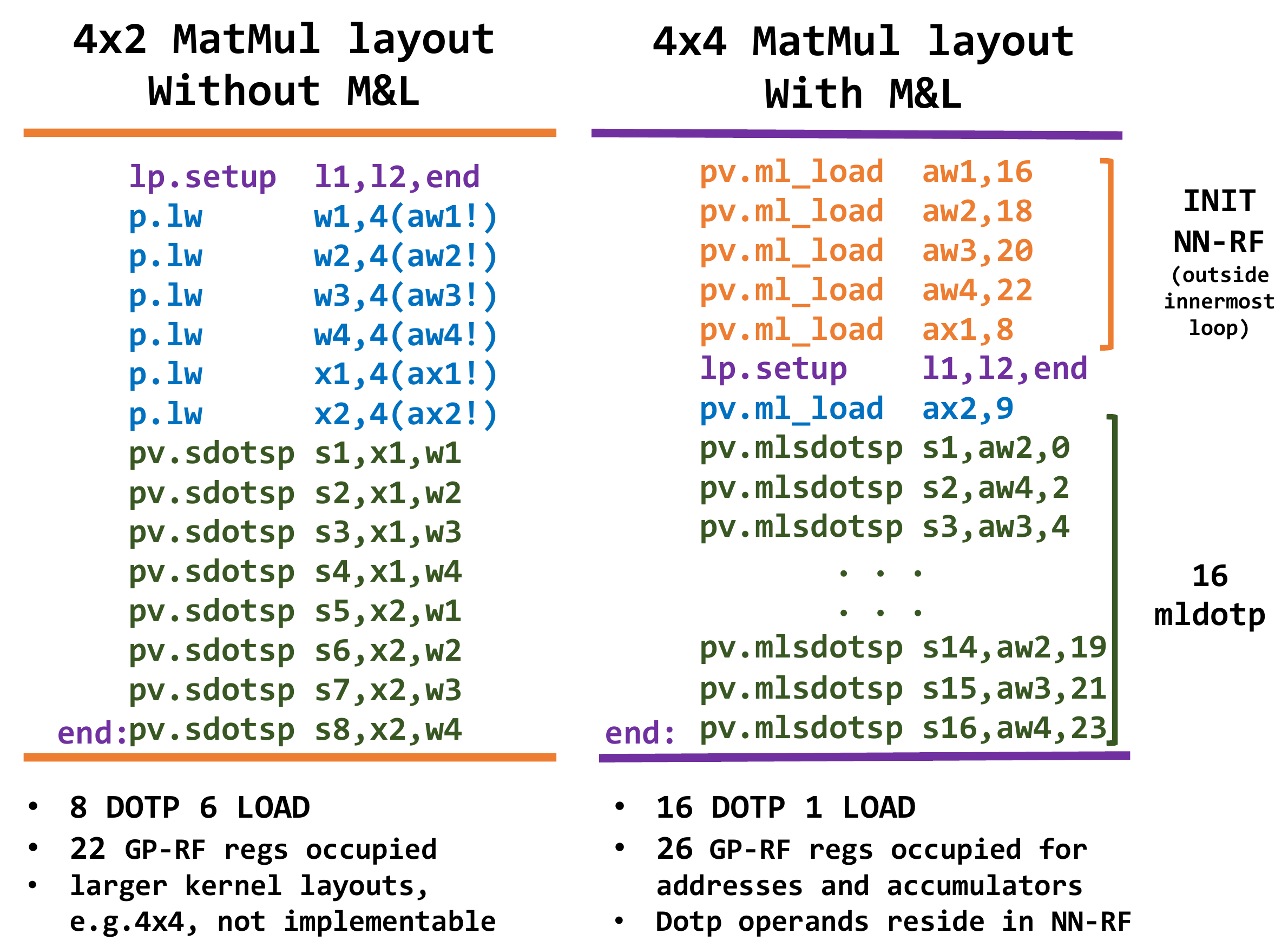}
         \label{fig:three sin x}
     \end{subfigure}
     \caption{a). Pipeline extension to the RI5CY core to support mixed-precision and M\&L instructions. b). Example of a \textit{MatMul} kernel using M\&L instruction, compared to the same kernel implemented without M\&L instruction. Thanks to the M\&L operating on the dedicated NN-RF, we can implement larger layouts of \textit{MatMul} kernels (right-sided) with a significant gain in terms of throughput. }
     \label{fig:riscy_nn}
\end{figure*}


The \textsc{Darkside} cluster's core, namely RVNN, is a 4-stage in-order single-issue pipeline, depicted in Fig.~\ref{fig:riscy_nn}a, that implements the RV32IMCF RISC-V \ac{ISA}, plus custom mixed-precision SIMD instructions operating on vector elements with power-of-two precision formats from 2-bit to 32-bit and all their possible permutations, supported through a dynamic bit-scalable execution~\cite{ottavi2020mixed}: the  instruction encoded into the \ac{ISA} identifies only the type of SIMD operations to be performed (denoted as \textit{Virtual Instruction}), while its format (i.e. the precision of the operands) is specified at run-time by reading the content of a specific \textit{Control\&Status Register} (CSR) of the core, which is writable by the programmer to set the desired precision, including mixed-formats. The SIMD instructions include \textit{dot-product (dotp)} based operations relevant to speed-up low-bitwidth compute-intensive kernels like Matrix-Matrix and Matrix-Vector multiplications. 

The micro-architecture of RVNN  is built on the baseline of the RI5CY~\cite{gautschi2017near} core and is reported in Fig.~\ref{fig:riscy_nn}a: we extend the ALU and the \textit{Dot-product} Unit to process 2-bit and 4-bit SIMD operations which are not supported by RI5CY, we add extra CSR registers to store the instruction formats' information and we integrate the \textit{Mixed-Precision Controller} (MPC) into the ID-STAGE of the pipeline. When a mixed-precision \textit{dotp} SIMD operation is performed, the decoder issues the \textit{Virtual Instruction} to select the specific compute unit to be used in the EX-STAGE of the pipeline, the format of the operands is specified by the CSR, while other control signals required for the execution are provided by the MPC. The \textit{Dot-Product} Unit, as shown in Fig.~\ref{fig:riscy_nn}a, is preceeded by a \textit{Slicer and Router} network, controlled by the MPC, which slices the registers according to the format (FMT) specified by the MPC; it selects the sub-portion of the vector RS2 to be used in the current operation and sign-extends (or zero-extends) it to match the size of the vector in RS1; afterwards, the network routes the operands to the appropriate set of multipliers. To minimize the logic necessary to implement the new extensions, the first operand of the mixed-precision operations (RS1) is designated to be always the highest precision operand, without loss of generality given the commutative property of add and multiply operators. The extended pipeline entails 17\% area overhead and 3\% power overhead compared to RI5CY, but it improves the performance on sub-byte and mixed-precision kernels by a significant factor (up to 7.7$\times$). 

The key enhancement of RVNN is a fused MAC-load (M\&L) operation that applies to any mixed-precision SIMD \textit{dotp} instruction supported. The design of the M\&L collapses the SIMD MAC and the load operations into a single one-cycle latency instruction since the datapath activated by the MAC operation would not interfere with the Load-Store Unit of the processor, and the two units can run in parallel. Fig.~\ref{fig:riscy_nn}a shows the micro-architectural modifications to the cores' datapath to enable the M\&L. When the M\&L is executed, the two operands for the \textit{dotp} operation are fetched from a dedicated register file, namely the Neural Network Register File (NN-RF), and routed to the \textit{Dotp}-Unit through a multiplexer controlled by the MPC. At the same time, the accumulators reside in the GP-RF. The NN-RF consists of 6 32-bit registers and is sized to maximize the innermost loops performance of the PULP-NN~\cite{garofalo2020pulp} convolution routines, dedicating 4 out of 6 registers to layer's weights and 2 out of 6 registers to input activations. As visible in Fig.~\ref{fig:riscy_nn}a, this choice constraints the activations of the convolution layers always to feature higher precision than the weights in mixed-precision operations, which however is the common case in current state-of-the-art software solutions to deploy \ac{DNN} models at the extreme-edge of the IoT~\cite{rakka2022mixed}.

Since the M\&L operates on the NN-RF, the occupancy of the 32 32-bit registers of the \ac{GP-RF} is reduced by a significant factor, since it would only host the accumulators of the \textit{dotp} operations and the addresses for the memory accesses. As a consequence, we can implement compute kernels with a higher amount of data reuse without incurring overheads to move data back and forth from the stack in the innermost hot loop (Fig.~\ref{fig:riscy_nn}, right-sided kernel). This solution guarantees up to 1.7$\times$ performance improvements over the execution of the same kernel without M\&L, with an extra area overhead of only 5\%, necessary to integrate the NN-RF in the \textsc{EX-STAGE} of the core pipeline. When a M\&L instruction is executed, one of the two source operands from the \ac{NN-RF} can eventually be updated with new data fetched from memory by the LSU,  extended to operate on the \ac{NN-RF} with negligible area overhead. However, the data stored in the \ac{NN-RF} registers can be kept until necessary to allow a higher degree of flexibility for data reuse strategies: a second M\&L instruction encoded into the \ac{ISA} performs only the \textit{dotp} branch, with no register update. 

From an instruction count perspective, the M\&L brings significant advantages, as shown in Fig.\ref{fig:riscy_nn}b. After out-of-the-loop initialization of the dedicated Neural \ac{NN-RF}, we perform 16 SIMD \textit{dotp}-like operations and only a single explicit load (with no concurrent MAC) instruction. Therefore, we reduce the number of pure load instructions in the innermost loop of the kernel by a factor of 6, at the same time doubling the throughput, with an overall dot-product/cycle improvement of {{57\%}} compared to the same core not featuring the M\&L. On the contrary, the impact of the M\&L on the Performance, Power, and Area (PPA) metrics of the RVNN core is minimal. Overall, the M\&L implies a gate count increase of just {{8.3\%}}, without deteriorating the critical path of the core and with negligible power overhead. On the other hand, the core enhanced with the M\&L achieves up to {{94\%}} dot-product unit utilization, compared with {{58\%}} of the RI5CY baseline.

\begin{figure}
    \centering
    \includegraphics[width=0.95\linewidth]{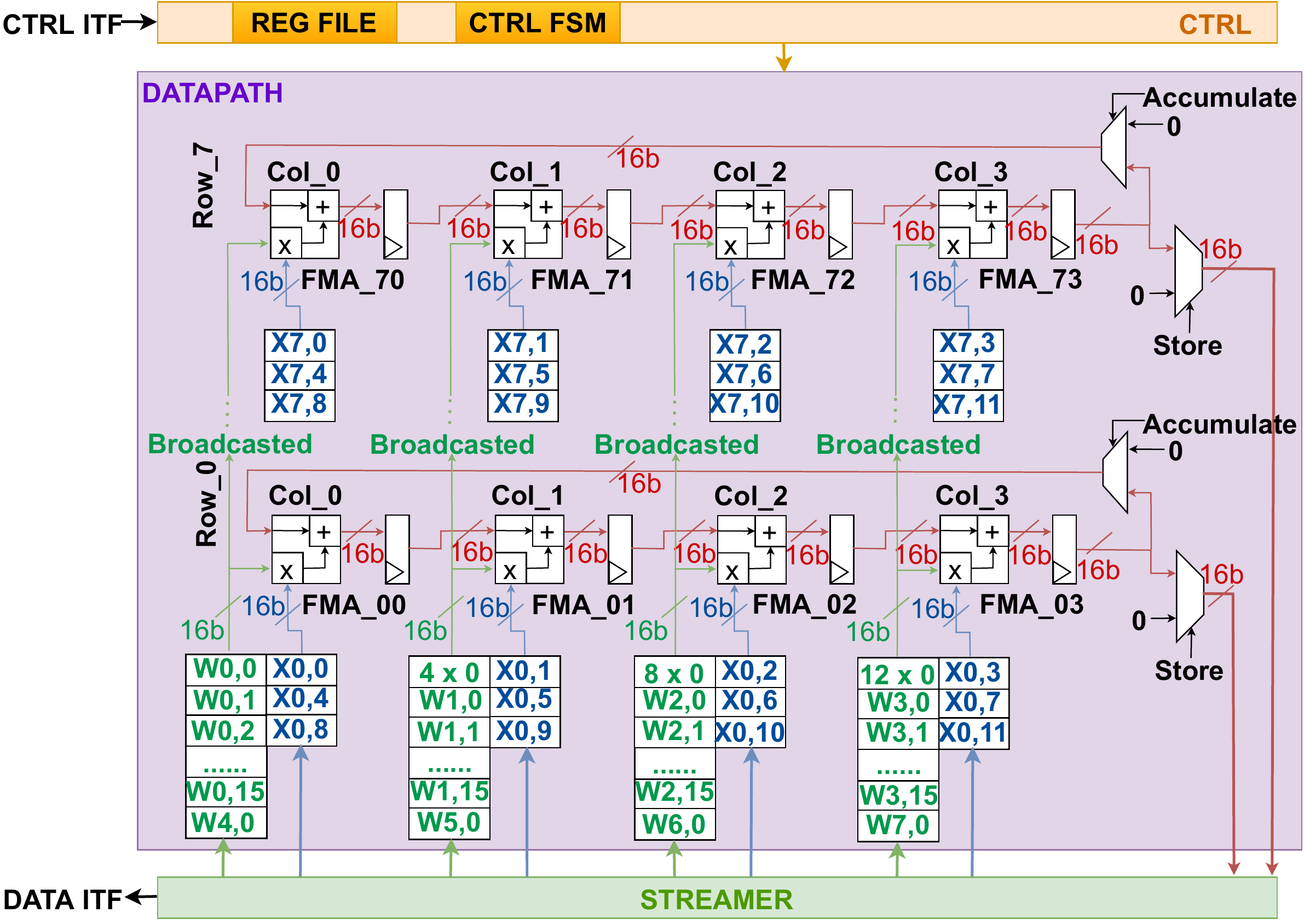}
    \caption{Architecture of the \ac{TPE}, with a focus on the interconnection  of the FMAs within the datapath.}
    \label{fig:tpe_detail}
\end{figure}

\subsection{Tensor Product Engine}
\label{sec:TPE_archi}
The Tensor Product Engine (\ac{TPE})~\cite{redmuledate2022} accelerates matrix multiplications (MatMuls) of the kind $ Z\;= X \cdot W$. It is designed to use the IEEE 754 binary-16 representation (FP16 in the following) since it is understood that \ac{FP}16 can be used to train Neural Networks without significant accuracy loss, but reducing the power consumption and time to computation \cite{nvidia_mixp}.
Fig.~\ref{fig:tpe_detail} shows the implementation of the \ac{TPE}. The datapath consists of 32 \ac{FP}16 \ac{FMA} units~\cite{tagliavini2018transprecision} organized in 8 rows, each of 4 columns. This configuration guarantees at the same time a good speed-up (from $9\times$ up to $22\times$) with respect to the software parallel execution, and an area overhead bounded to 44.8\% of the area occupied by all the RV-NN cores.
The \acp{FMA} along each row are cascaded; each \ac{FMA} passes the intermediate result as input to the unit to its right. 

To internally handle the computation of matrices larger than the array size, and to avoid intermediate store operations, the \acp{FMA} in each row are closed in a feedback loop so that the right-most \ac{FMA} feeds back the computed partial product as accumulation input to the left-most \ac{FMA} of the same row. Using this approach, the \ac{TPE} can exploit maximum reuse of both the \textbf{X}-matrix elements and the intermediate product, so that it stores the computed sub-blocks of the \textbf{Z}-matrix to the memory only at the very end of their computation.
To match the critical path of the cores, each \ac{FMA} features three internal pipeline registers. To maximize throughput, the \textbf{X}-matrix elements of each \ac{FMA} are held steady for the number of cycles necessary to the \acp{FMA} of each row to compute the partial results. On the other hand, \textbf{W}-matrix operands are streamed-in at each cycle and broadcasted to all the \acp{FMA} of the same column.
The memory accesses are scheduled so the load and store phases do not introduce overhead during the computation. This way, the \ac{TPE} can reach an overall 98.8\% utilization of the internal \acp{FMA} with a near-to-ideal performance (31.6 out of 32 MAC/cycle).
The computed sub-blocks of the \textbf{Z}-matrix are stored in the memory only at the end of their computation, maximizing internal data reuse. 

The \ac{TPE}, as well as the other accelerators integrated in \textsc{Darkside}, features a non-blocking event-based execution mode: the cores of the cluster, after programming the accelerator and starting its execution, can either go in sleep mode or resume software code execution. This mechanism enables complex execution models where the accelerator can be used in parallel with the general-purpose cores to boost the performance of the target kernel. This scenario is made possible also thanks to the dynamic arbitration mechanism provided by the \ac{HCI}, which allows the requests to the memory by the \ac{TPE} and the cores to be served simultaneously if there are no bank conflicts.

\subsection{Depth-Wise Convolution Engine}
\label{sec:DWE_archi}

\begin{figure}
    \centering
    \includegraphics[width=0.95\linewidth]{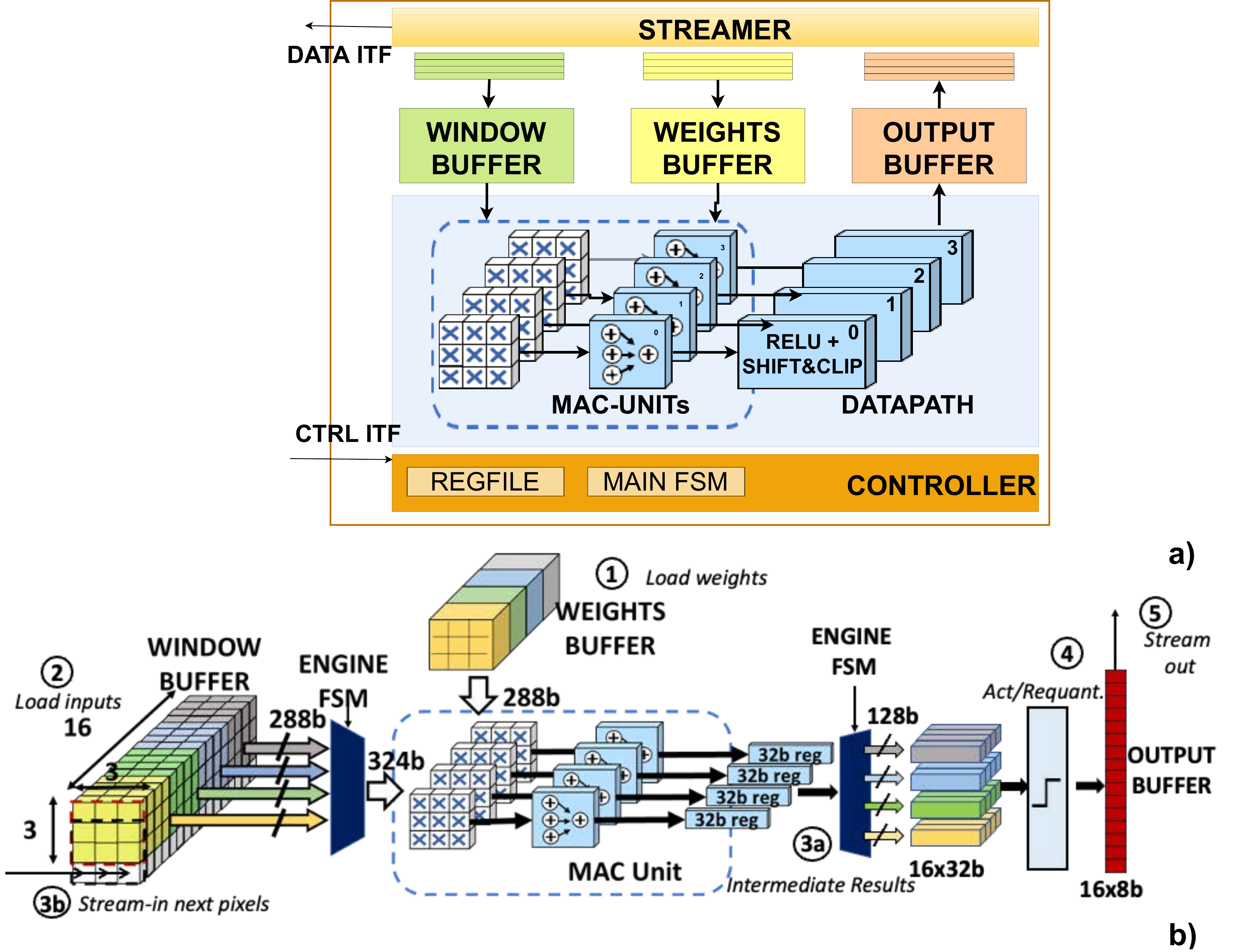}
    \caption{(a) Architecture overview of the Depth-wise digital accelerator, enclosed in the HWPE. (b) Execution flow of the depth-wise operation.}
    \label{fig:dwe_archi}
\end{figure}

The Depth-Wise Convolution Engine (\ac{DWE}) can process the low-reuse depth-wise component of the depth-wise + point-wise kernels often used in recent neural network models for mobile applications, leaving the much better-parallelizable point-wise kernels to the M\&L-accelerated software. The \ac{DWE} processes 8-bit signed input and weight tensors stored in L1 memory using the \ac{HWC} layout, the same used by the cores to execute the point-wise kernels and therefore requiring no time-consuming intermediate on-the-fly marshalling operations. The 8-bit output tensors are generated after applying re-quantization steps.

The architecture is shown in Fig.~\ref{fig:dwe_archi}a. It employs a weight-stationary data flow to maximize the data reuse and targets 3$\times$3 depth-wise layers, the one most commonly encountered in \acp{DNN}. Although its datapath is optimized for 3$\times$3 depth-wise convolutions, the DWE can be used to run kernels with different sizes using the same approach presented in~\cite{meloni2018neuraghe}, at the cost of additional data manipulation on the intermediate results, hence less computation efficiency.

The execution flow is depicted in Fig.~\ref{fig:dwe_archi}b. The weights from 16 3$\times$3 filters are loaded into the \textit{Weights Buffer} \circled{1}, before the execution starts. The weights are kept in the buffer until they have been used to scan the whole input tensor. 
The input image is filtered through a vertical sliding window on the spatial dimensions, using a \textit{window buffer} of 4$\times$3$\times$16 registers. The first three rows are loaded at the beginning of the iteration \circled{2} and consumed in 4 cycles by the datapath of the \ac{DWE} consisting of 36 MAC units. The intermediate results are accumulated over 16 32-bit buffers, accessed 4 at a time in the 4-cycle operation loop \circled{3a}. Afterwards, non-linear activation functions and ancillary operations such as shifting and clipping are applied to re-quantize the results to 8-bit precision \circled{4}. After 4 cycles of operation, the 16-channel 8-bit pixels stored in the \textit{output buffer} are streamed out of the accelerator \circled{5}. Meanwhile, the streamer uses three cycles (overlapped with the computation) to fill the fourth row of the \textit{window buffer} \circled{3b}, needed to implement the sliding window mechanism.

The \ac{DWE} is designed to keep the datapath always active, in all stages and to fully exploit the memory bandwidth of 36B per cycle available on the cluster, achieving the overall performance of 30 MAC/cycle, more than 10$\times$ better than a software execution of the depth-wise kernels, on the 8 RVNN cores. 

\subsection{DataMover}
\label{sec:datamover}
\begin{figure}[t]
    \centering
    \includegraphics[width=0.9\linewidth]{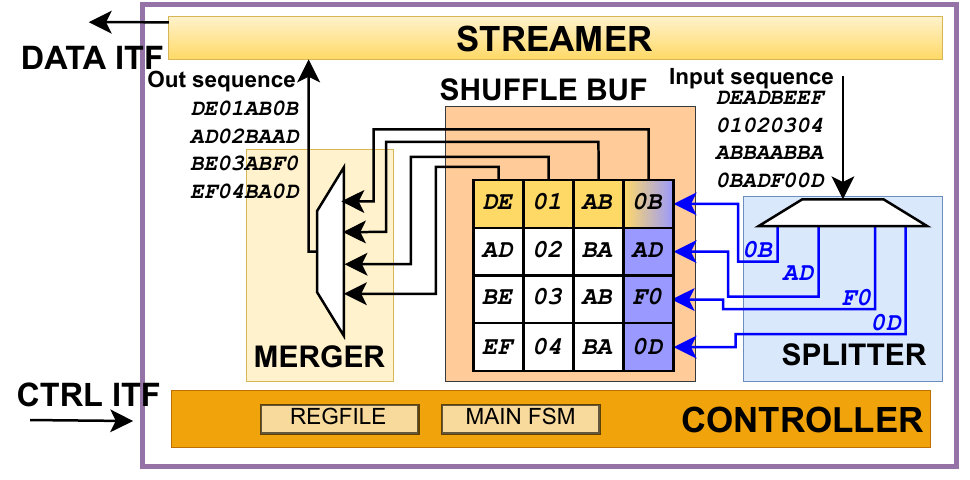}
    \caption{Overview of the DataMover architecture and execution flow.}
    \label{fig:datamover}
\end{figure}
On-the-fly data marshalling operations can dramatically reduce the performance of \ac{DNN} workloads, during both inference and training tasks. To perform efficiently on-the-fly data transposition, the \textsc{Darkside}'s cluster is enhanced with a DataMover unit, exposed to the \ac{HCI} with an additional master port on the \textit{logarithmic} branch. The architecture is depicted in Fig.~\ref{fig:datamover}. It consists of a tiny accelerator of only 54 kGE, capable of transposing \ac{DNN} 3-dimensional tensors stored in the L1 memory, with 1.5-100$\times$ less time than eight RVNN cores and increased energy efficiency up to 50$\times$ (the lower the precision of chunks to transpose the more significant the advantages).

The accelerator works on data with configurable precision, $d$, in the range from 32-bit down to 1-bit. It splits incoming data streams from memory into chunks of size $d$, internally buffered into the \textit{Shuffle Buffer} which features 32$\times$ 32-bit register with the transposed output format; only 32/$d$ of them are used, depending on the data size configuration $d$. After 32/$d$ input stream transactions, the transposed words are streamed out to the L1 memory and the accelerator continues the operations with new input streams. 

\section{Chip Implementation and Measurements}

\begin{figure}[t]
    \centering
    \includegraphics[width=0.95\linewidth]{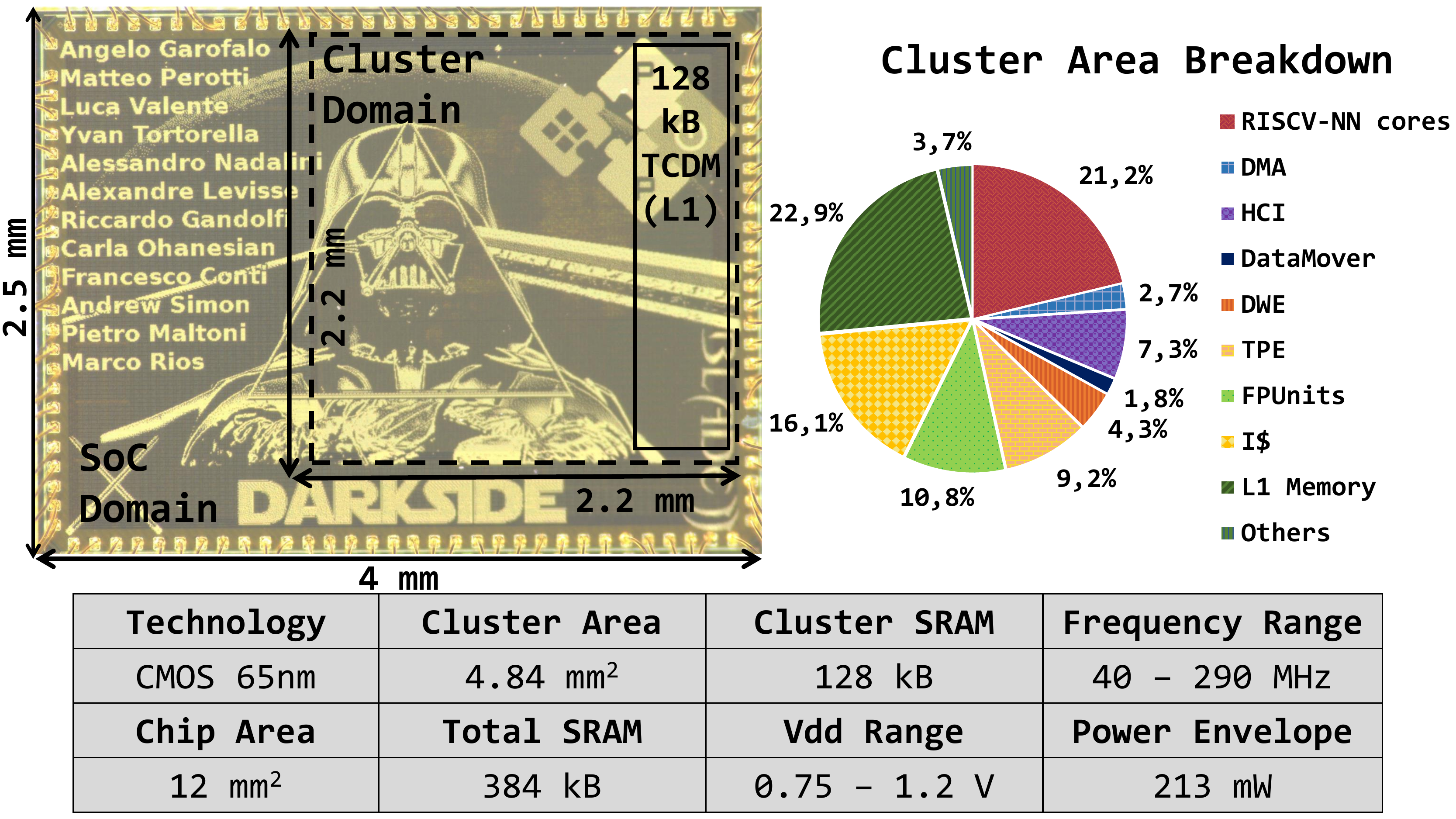}
    \caption{Chip micrograph and specifications. Area breakdown of the cluster.}
    \label{fig:die_micrograph}
\end{figure}

\begin{figure}[t]
    \centering
    \includegraphics[width=0.95\linewidth]{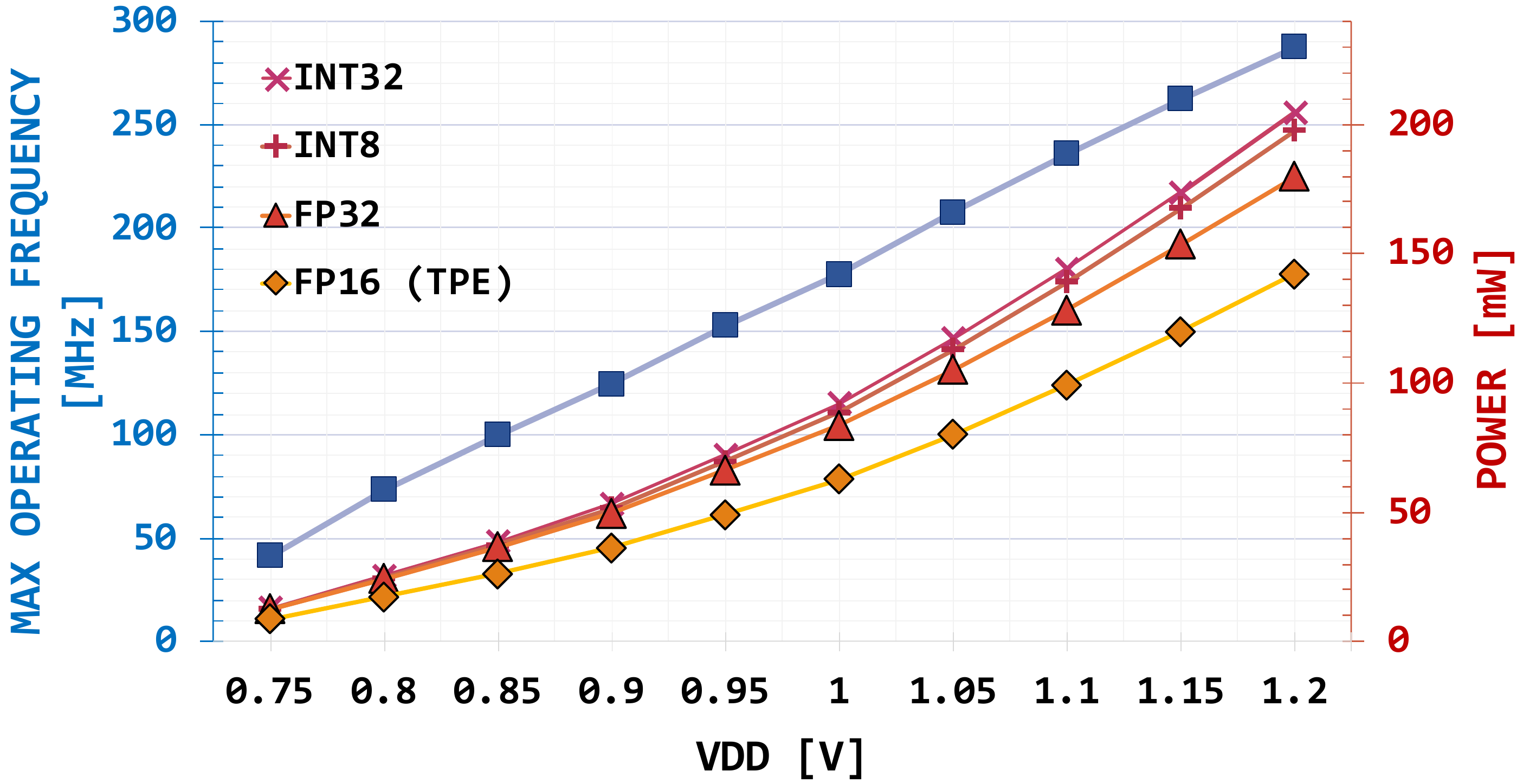}
    \caption{Voltage sweep vs. max frequency vs. power consumption.}
    \label{fig:sweep}
\end{figure}

\begin{figure}[t]
    \centering
    \includegraphics[width=0.95\linewidth]{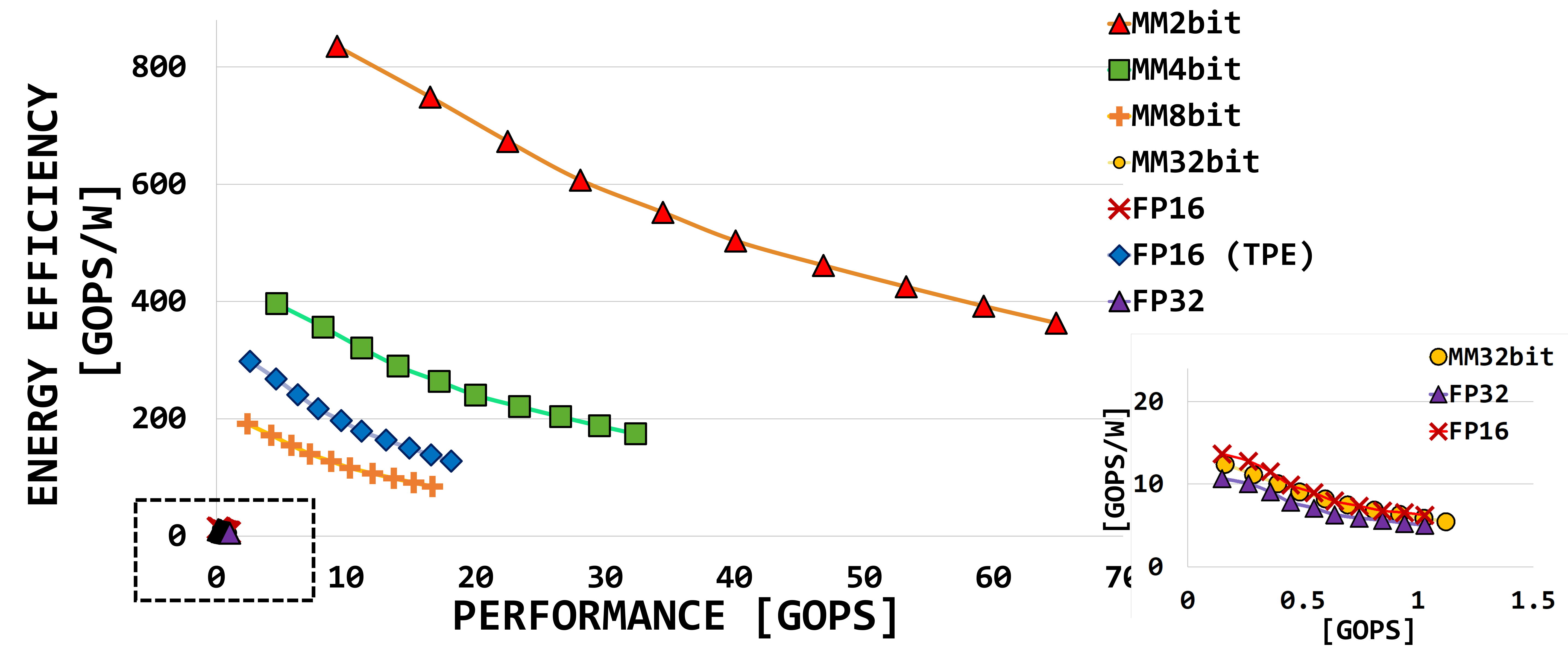}
    \caption{Performance and Energy Efficiency of the \textsc{Darkside} Cluster. The measurements are performed at the max frequency, sweeping the supply voltage between 0.75V and 1.2V.}
    \label{fig:gops_vs_gopsw}
\end{figure}

Fig.~\ref{fig:die_micrograph} shows the chip micrograph of \textsc{Darkside}. The SoC is implemented with TSMC 65nm technology, targeting a clock frequency of 250 MHz in worst-case operating conditions. The die area, including the \textit{Fabric} domain, is 12 mm$^2$, while the cluster area is 4.84 mm$^2$, partitioned as shown in Fig.~\ref{fig:die_micrograph}. The majority of the cluster's area is occupied by cores,  hierarchical I\$ and 128kB of L1 memory, while the accelerators account only for 15.3\% of the total area. 

The measurements of the \textsc{Darkside}'s cluster are performed using an Advantest SoC hp9300 integrated circuit testing device, which precisely regulates the supply voltages delivered to the SoC and allows accurate current measurements of the SoC's cluster power domain. 
Fig.~\ref{fig:sweep} reports the maximum operating frequency and the power consumption of the cluster over the 0.75V to 1.2V voltage range. The operating frequency increases linearly with the supply voltage up to 290 MHz at 1.2V. The power is measured on the silicon prototype, running integer and floating-point compute-intensive kernels (MatMuls). Offloading \ac{FP}16 MatMuls on the \ac{TPE} saves 30\% of the power compared to the execution on the 8 cores. 

Fig.~\ref{fig:gops_vs_gopsw} shows the cluster's performance and efficiency on integer and floating-point MatMuls, sweeping all the main supported data formats. The measurements are taken at the maximum operating frequency, sweeping the supply voltage from 0.75V to 1.2V. On ML workloads (i.e., integer 8-/4-/2-bits) deployed on RV-NN cores, \textsc{Darkside} delivers up to 65 GOPS with a peak efficiency of 835 GOPS/W. The TPE boosts the performance and the efficiency of FP16 MatMuls to 18.2 GFLOPS and 300 GFLOPS/W, respectively, 17.7$\times$ and 21.8$\times$ higher compared to a software scalar execution on the 4 FPUs of the \textsc{Darkside} cluster.

\section{Benchmarking}
To demonstrate the capabilities of \textsc{Darkside} on \ac{SoA} DNN workloads, we benchmark mixed-precision convolution kernels, end-to-end inference of the MobileNetV2 and one TinyML training use-case.

\subsection{Single DNN Kernels}

\begin{figure}[t]
    \centering
    \includegraphics[width=\linewidth]{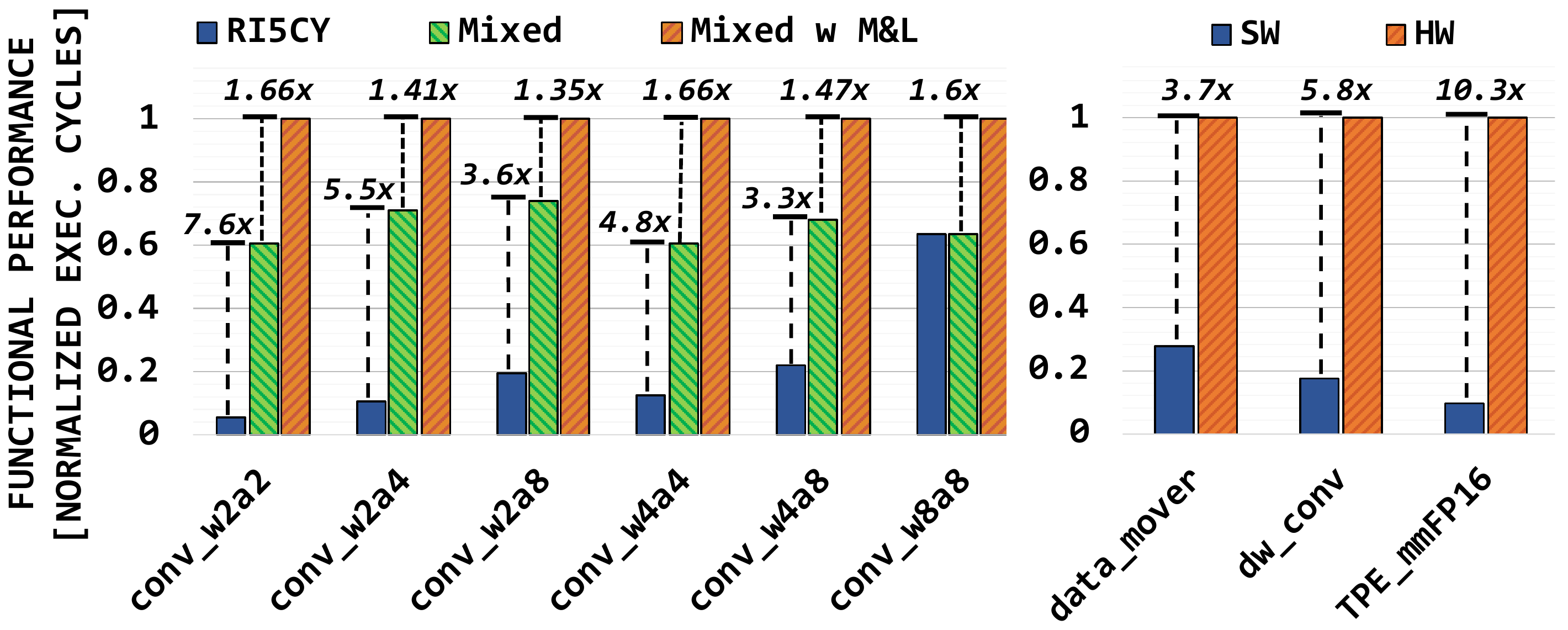}
    \caption{Left: Normalized single-core performance of mixed-precision convolutions. RVNN core (\textit{Mixed w M\&L}) is compared against a core featuring only mixed-precision \textit{dotp} operations but no M\&L (\textit{Mixed}) and against \textit{RI5CY}, which features no extensions for quantized neural networks. Right: performance improvement of execution on cluster-coupled accelerators over software (8-cores).}
    \label{fig:normalized_perf}
\end{figure}

To highlight the features of the RVNN cores, we show improvements over the baseline, benchmarking several convolution kernels. To measure the computing performance, the layers operate on data stored in the L1 memory, with 64 3$\times$3$\times$32 filters applied on a 32$\times$16$\times$16 input feature map, spanning different integer data formats (from 8- down to 2-bit) for the inputs and the weights, including mixed-precision cases. Results are reported in Fig.~\ref{fig:normalized_perf} in terms of normalized execution cycles (with respect to the RVNN core with M\&L). As shown, the M\&L instruction improves the performance by up to 1.7$\times$ compared the execution with baseline mixed-precision SIMD \textit{dotp} and load instructions (\textit{Mixed} in the figure). Overall, ISA and micro-architecture design of RVNN leads to a cumulative performance improvement of up to 13$\times$ with respect to RI5CY, which supports only 8-bit SIMD operations and no M\&L mechanisms. 

Analyzing the depth-wise kernels, in Fig.~\ref{fig:normalized_perf} we show that this workload achieves at least 5.8$\times$ better performance by offloading it to the dedicated digital accelerator presented in this work, the Depth-wise engine (\textit{DWE}), instead of running it on 8 RVNN cores. This conclusion is also strengthened in Sec.~\ref{sec:end_to_end_inference} on a real-life \textit{Bottleneck} layer use-case. Furthermore, we show that the DataMover can reduce by more than 3.7$\times$ execution cycles on 8-bit data marshalling operations, compared to the same task offloaded to the 8 cores.

On 16-bit floating-point (FP16) matrix-multiplication workloads, the TPE boosts the performance by up to 10.3$\times$ with respect to the software execution of the same kernels exploiting the FP16 SIMD instructions available on the 4 floating-point units (FPUs) present in the cluster~\cite{tagliavini2018transprecision}. 

\subsection{End-to-End MobileNetV2}
\label{sec:end_to_end_inference}

%
First, we present the results of benchmarking the \textit{Bottleneck} layer, the core building block of the MobileNetV2. We demonstrate our improvements incrementally by comparing our architectural solutions over a reference cluster that features 8 RI5CY cores (without the mixed-precision SIMD operations and the M\&L custom ISA extensions proposed in this work) and no dedicated accelerator.
To implement the software to execute the \textit{Bottleneck} we use the PULP-NN library (which we use as-is to benchmark the reference cluster), extended to include additional kernels to exploit the new ISA instructions implemented in the RV-NN cores and a set of hardware-abstraction-layer (HAL) functions to program and start the accelerators that the programmer can easily insert into the C code. We adopt the 8-bit signed integer representation for all the tensors of the \textit{Bottleneck}.

\begin{figure}
    \centering
    \includegraphics[width=0.75\linewidth]{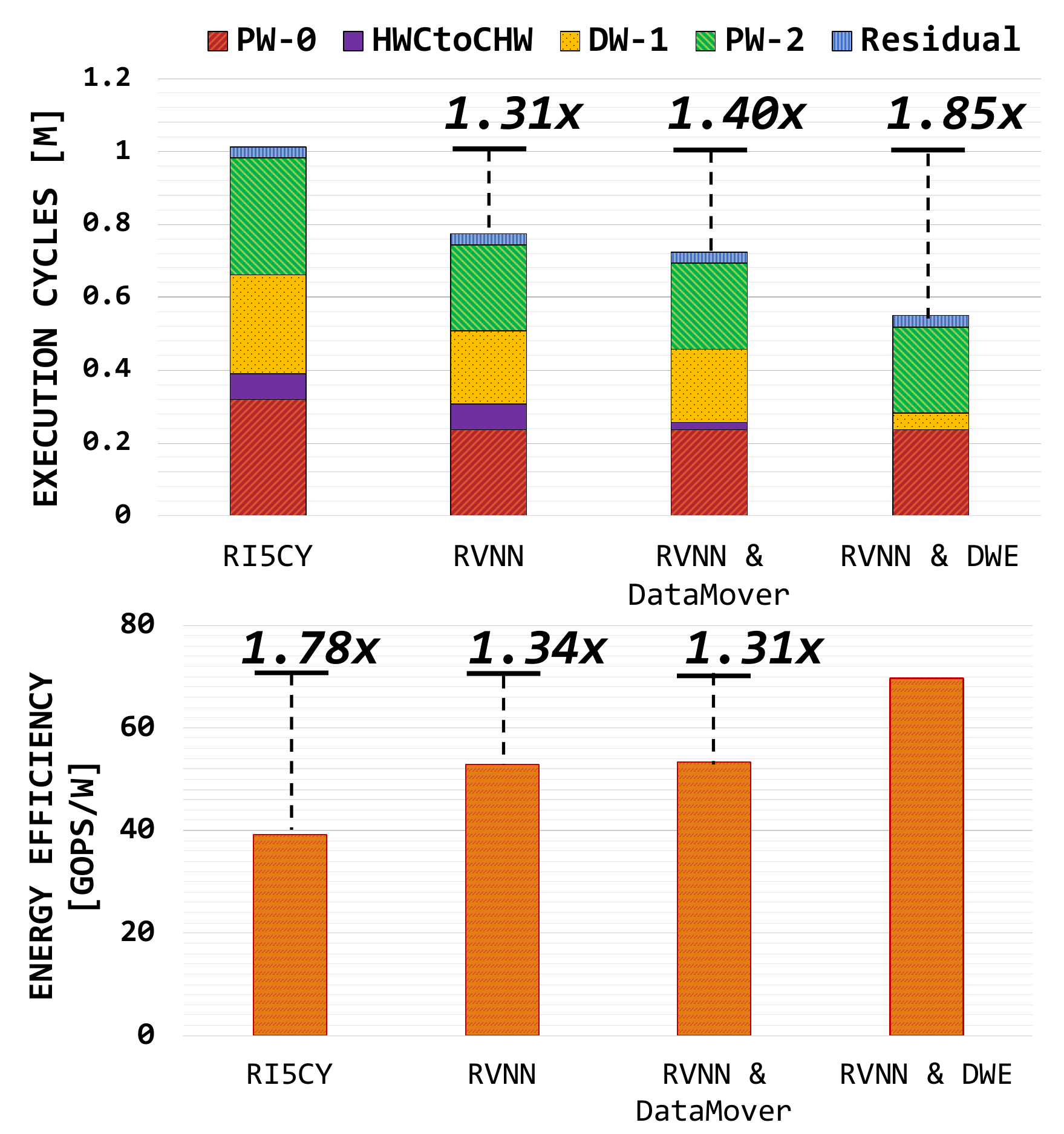}
    \caption{Comparison of the \textit{Bottleneck} layer execution, offloaded to different hardware compute units. a) reports the execution latency (in terms of compute cycles), b) reports the energy efficiency (in GOPS/W). The cluster is running at the best performance operating point: $f_{CLU}\,=\,290\,MHz$, $V_{DD}\,=\,1.2\,V$.}
    \label{fig:bottleneck_results} 
\end{figure}

The results, in terms of execution cycles and energy efficiency, are reported in Fig.~\ref{fig:bottleneck_results}. The M\&L improves the execution of point-wise and depth-wise layers on 8 RVNN cores by 1.31$\times$ compared to the execution on 8 RI5CY cores. Additional 1.13$\times$ improvements are given by the data transposition (i.e. HWC to CHW data marshalling) performed by the DataMover, instead of transposing data via software. Finally, the DWE allows to speed-up the execution of depth-wise convolution by 4.4$\times$ compared to the execution on 8 RV-NN cores, with a final performance improvement of 1.85$\times$ on the whole \textit{Bottleneck} layer compared to the RI5CY baseline.

To put the previous results in perspective, we benchmark \textsc{Darkside} on the end-to-end inference task of the MobileNetV2 model. We employed the standard MobileNetV2 with depth multiplier 1.0 and input size 224$\times$224, composed of 16 stacked \textit{Bottleneck} layers. The input and weight tensors feature 8-bit precision for all the depth-wise and Conv2d layers. The point-wise layers feature 8-bit input tensors, while the weights are represented with reduced precision, i.e. 4-bits. This mixed-precision configuration of the model achieves similar Top-1 accuracy (only a 2.5\% drop compared to the 8-bit version) while ensuring almost 2$\times$ less memory footprint (1.07 MB) for the network weights~\cite{schaefer2022edge}. 

\begin{figure}[t]
    \centering
    \includegraphics[width=0.95\linewidth]{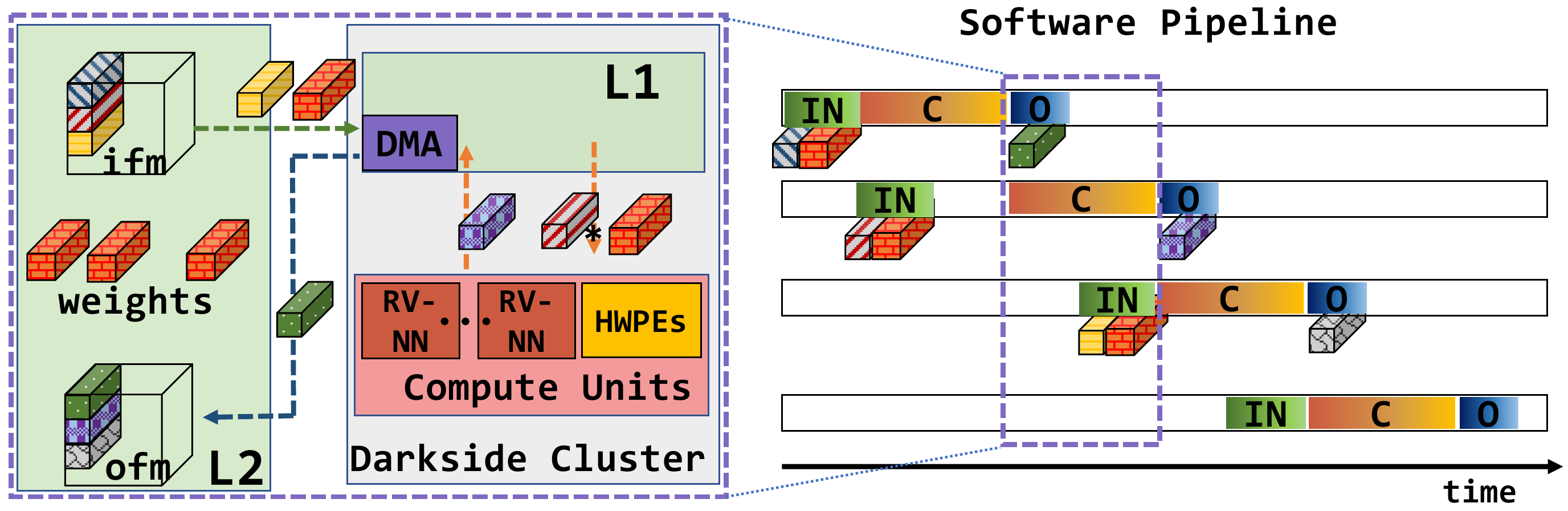}
    \caption{DNN tiling software pipeline. The figure shows the concurrent execution of weights and activations transfer (from  L2 to  L1 memory, indicated with \textbf{\textit{IN}}), the computation of the kernel (indicated with \textbf{\textit{C}}) and the copy of produced output tiles from  L1 to  L2 memory (indicated with \textbf{\textit{O}}).}
    \label{fig:sw_pipe}
\end{figure}

To enable the computation on the cluster, both  weights and  activations of the model must be divided into tiles that fit 128 kB of the L1 SRAM. Therefore, we assume the weights and the feature maps for all the network layers to be stored in the off-the-cluster L2 memory, and we adopt the data and execution flow presented in Dory~\cite{burrello2021dory}. Dory is used to calculating the data tiling solutions fitting the L1 memory constraints and to schedule the data transfers from L2 to L1 and vice-versa, performed through the cluster DMA in double-buffering. The described software pipeline is represented in Fig.~\ref{fig:sw_pipe}. For cases where the execution is not memory-bound, data movements overlap with the computation, with negligible overhead ($\leq 5\%$) to the execution latency.

%
However, since \textsc{Darkside}'s \textit{Fabric} domain has the only purpose of acting as a programmable testbench for the cluster, it features a small L2 memory which is insufficient to host the entire MobileNetV2. Therefore, to benchmark the computing capabilities of the cluster on real-life end-to-end DNN models, we exploit our previous experience on explicit memory management, data tiling techniques~\cite{burrello2021dory} and on the deployment of real-sized DNN models on application chips such as Vega~\cite{VEGA} to build a model of the system, with larger L2 memory, on which we run the experiments. The hardware-oriented description of the SoC is integrated into our open-source~\footnote{https://github.com/pulp-platform/gvsoc} event-based emulator, called GVSOC~\cite{bruschi2021gvsoc}; to run the experiments, the following measurements and considerations are taken: 
\begin{enumerate}
    \item We assume to have a L2 memory of 2MB, necessary to host the entire MobileNetV2 model and to store the program code;
    \item We analyze the traffic between L2 and L1 memories by running end-to-end simulations of the MobileNetV2 on the GVSOC; as expected, during the execution of the inference task we are never memory-bound; therefore, the contribution of the L2 to L1 (and vice-versa) data movements is relevant only for the total energy consumption;
    \item We conduct silicon measurements, in terms of latency and energy, on all the L2 to L1 data transfers (and vice-versa) necessary to compute each tile and determined by the GVSOC simulations; we then include the measurements in the model;
    \item We conduct silicon measurements, in terms of latency and the energy, on all the kernels necessary to compute each tile generated by the Dory framework; we then include the measurements in the model.
\end{enumerate}


\begin{figure}[t]
    \centering
    \includegraphics[width=\linewidth]{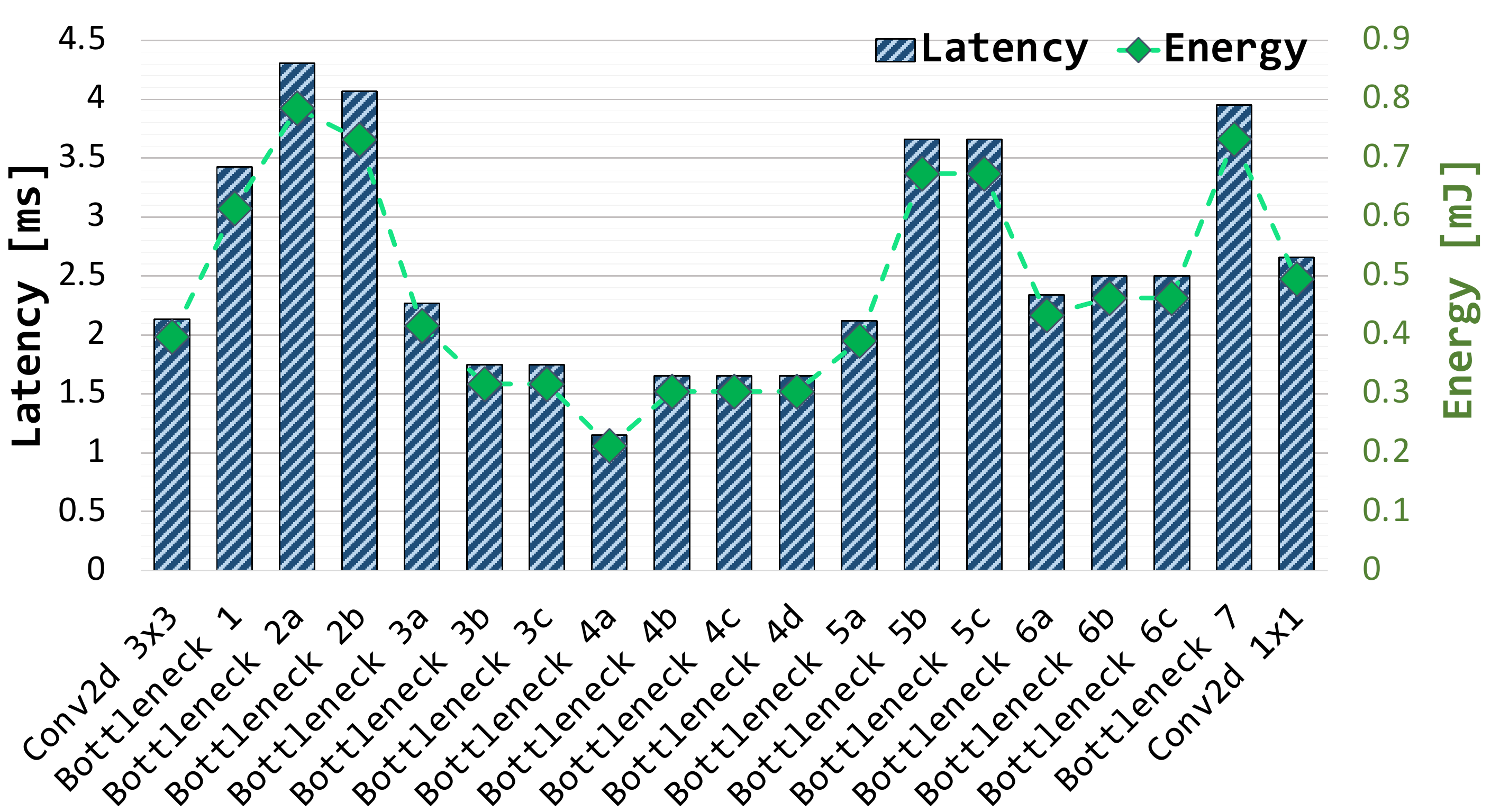}
    \caption{Layer-wise execution latency and energy of the MobileNetV2 on \textsc{Darkside}'s cluster running at $f_{clk}=290\,MHz$ and $V_{DD}=1.2\,V$.}
    \label{fig:mbntv2}
\end{figure}
The layer-wise compute time and energy of the inference task are shown in Figure~\ref{fig:mbntv2}. \textsc{Darkside} can perform the entire end-to-end task with a performance of more than 20 frame/s, with an energy budget of 11mJ. The performance is 2$\times$ better than the one achieved on the Vega's cluster running at 250 MHz~\cite{VEGA}, thanks to our architectural contributions, the M\&L extensions that accelerate the point-wise kernels and the dedicated DWE that boosts the execution of depth-wise convolutions. Despite Vega being implemented in the 22nm technology node, our end-to-end energy consumption of 11 mJ remains still comparable, in the same order of magnitude.


\subsection{TinyML On-Chip Training}
The \ac{TPE} enhances the \textsc{Darkside} cluster to support efficient \ac{FP} matrix-matrix multiplications, enabling de-facto on-chip TinyML training workloads. To benchmark the \ac{SoC} in terms of execution latency and energy on real-sized problems, we execute the \ac{AE} \ac{DNN} model~\cite{unknown}, commonly used within the TinyML scenario for unsupervised anomaly detection tasks. The TinyML \ac{AE} consists of Encoder and Decoder layers (made by 128 unit Fully Connected layers with BatchNorm and ReLu activation functions) and a latent space layer of size 8. The input and the output size is 640. We benchmark the whole training stage (forward and backward steps within one training \textit{epoch}), adopting a batch size of 16, which is a reasonable trade-off between performance and memory occupation for \ac{IoT} multi-core microcontroller-class devices. The tensors are represented with the \ac{FP}16 format, and we adopt the same data flow explained above, which uses tilings and double-buffering.

To highlight the boost given by the \ac{TPE} and the DataMover, we first implement the \ac{AE} on the 8 general-purpose RVNN cores (we call this configuration \textit{SW}), which share 4 floating-point units supporting \ac{FP}16 formats, using a software library optimized for on-chip training~\cite{nando2022samos}. Then, we implement the \ac{AE} offloading the matrix-multiplication workload to the \ac{TPE} (\textit{TPE} configuration), still performing on the cores control tasks (e.g. programming the DMA for double buffering, programming the \ac{TPE} control units) and matrix transpositions. As a third execution mapping, we use the DataMover to speed-up also the matrix transpositions (\textit{TPE + DataMover}). 

\begin{figure}[t]
    \centering
    \includegraphics[width=0.95\linewidth]{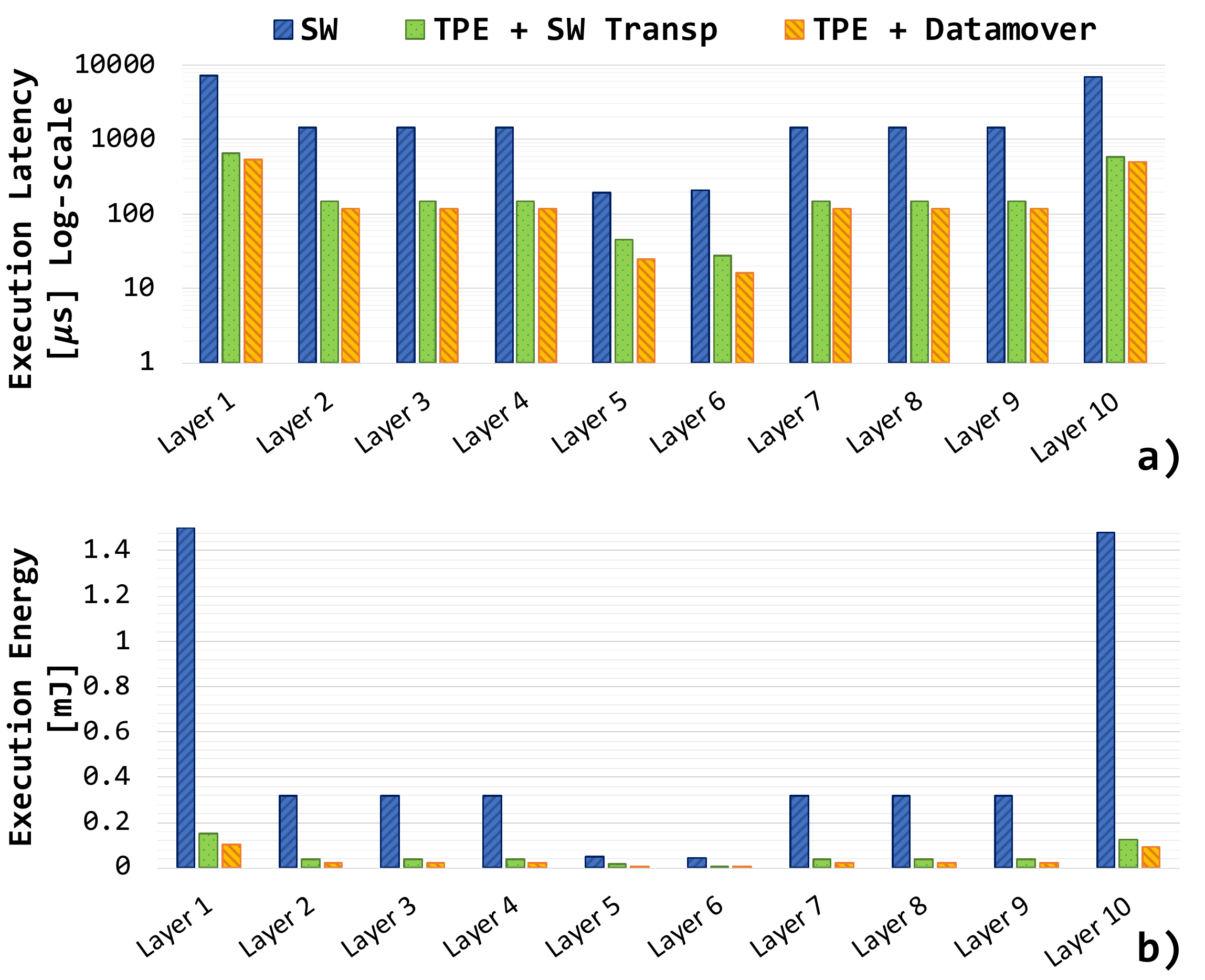}
    \caption{a) Layer-wise execution time of the AutoEncoder (AE) TinyML model. b) Execution Energy of the AE, including the energy for L2 to L1 (and vice-versa) data movements. Cluster running at 290MHz and 1.2.}
    \label{fig:ae_results}
\end{figure}
The results are reported in Fig. \ref{fig:ae_results} in terms of execution latency and energy consumption. As expected, the \ac{TPE} delivers at least 10$\times$ speed-up with respect to a pure SW execution on all the layers of the \ac{AE} except for the latent space layers, where the performance improvement is reduced to 4-7 $\times$ due to lower arithmetic intensity of those layers. The matrix transposition performed with the tiny DataMover accelerator contributes to an additional 1-2 $\times$ of speed-up. Overall, combining the \ac{TPE} and the DataMover, the entire training epoch runs in 1.8 ms with an energy consumption of $345\,\mu J$, 13 $\times$ faster than the \textit{SW} execution of the \ac{AE} on the 8 RV-NN cores, with 14$\times$ lower energy consumption.

\section{Comparison with the State-of-the-Art}

\begin{table*}[t]
\caption{Comparison With State Of The Art.}
\label{tab:soa}
\centering
\resizebox{\textwidth}{!}{
     \begin{tabular}{c|c|c|c|c|c}
\textbf{}                                                                      & \textit{SleepRunner~\cite{bol2021sleeprunner}}                                            & \textit{SamurAI~\cite{miro2020samurai}}                                                 & \textit{VEGA~\cite{VEGA}}                                                               & \textit{Dustin~\cite{ottavi2022dustin}}                                                                                           & \textit{This work}                                                                                       \\ \hline
\textit{Technology}                                                            & \begin{tabular}[c]{@{}c@{}}CMOS 28nm\\      FD-SOI\end{tabular} & CMOS 28nm        FD-SOI                                          & \begin{tabular}[c]{@{}c@{}}CMOS 22nm\\      FD-SOI\end{tabular}             & CMOS 65nm                                                                                                 & CMOS 65nm                                                                                                \\ \hline
\textit{Die Area}                                                              & 0.68   mm2                                                      & 4.5   mm2                                                        & 12   mm2                                                                    & 10   mm2                                                                                                  & 12 mm2                                                                                                   \\ \hline
\textit{Application}                                                           & IoT   GP                                                        & IoT   GP + DNN                                                   & IoT   GP + NSAA+DNN                                                         & IoT   GP + DNN + QNNs                                                                                     & \begin{tabular}[c]{@{}c@{}}IoT GP + NSAA   \\      + DNN + QNNs\end{tabular}                             \\ \hline
\textit{CPU}                                                                   & CM0DS                                                           & 1x   RI5CY                                                       & 10   x RI5CY                                                                & \begin{tabular}[c]{@{}c@{}}16   x RI5CY \\      MPIC CORES\end{tabular}                                   & 8 x RI5CY-NN                                                                                             \\ \hline
\textit{ISA}                                                                   & Thumb-2 subset                                                  & RVC32IMFXpulp                                                    & RVC32IMFXpulp+S                                                             & RVC32IMCXmpic                                                                                             & RVC32IMFXpulpNN2                                                                                         \\ \hline
\textit{\begin{tabular}[c]{@{}c@{}}Int Precision\\ (bits)\end{tabular}}        & 32                                                              & 8,   16, 32                                                      & 8,   16, 32                                                                 & \begin{tabular}[c]{@{}c@{}}2,  4, 8, 16, 32\\ (plus   Mixed-Precision)\end{tabular}                       & 2,  4, 8, 16, 32    (plus mixed-precision)                                                               \\ \hline
\textit{FP   Precision}                                                        & --                                                              & --                                                               & FP32, FP16,   bfloat                                                        & --                                                                                                        & FP32, FP16                                                                                               \\ \hline
\textit{Supply   Voltage}                                                      & 0.4 - 0.8 V                                                     & 0.45 - 0.9 V                                                     & 0.5 – 0.8 V                                                                 & 0.8 - 1.2 V                                                                                               & 0.75 - 1.2 V                                                                                             \\ \hline
\textit{Max   Frequency}                                                       & 80 MHz                                                          & 350 MHz                                                          & 450 MHz                                                                     & 205 MHz                                                                                                   & 290 MHz                                                                                                  \\ \hline
\textit{Power Envelope}                                                        & 320   µW                                                        & 96   mW                                                          & 49.4   mW                                                                   & 156   mW                                                                                                  & 213 mW                                                                                                   \\ \hline
\textit{\begin{tabular}[c]{@{}c@{}}Best   Integer \\ Performance\end{tabular}} & 31 MOPS (32b)                                                   & 1.5 GOPS (8b)                                                    & 15.6 GOPS (8b)                                                              & \begin{tabular}[c]{@{}c@{}}15 GOPS (8b)   \\      30 GOPS (4b) \\      58 GOPS (2b)\end{tabular}          & \begin{tabular}[c]{@{}c@{}}17 GOPS (8b) \\      32 GOPS (4b) \\      65 GOPS (2b)\end{tabular}           \\ \hline
\textit{\begin{tabular}[c]{@{}c@{}}Best Integer\\ Efficiency\end{tabular}}     & 97   MOPS/mW (32b)                                              & \begin{tabular}[c]{@{}c@{}}230   GOPS/W\\      (8b)\end{tabular} & 614   GOPS/W  (8b)                                                          & \begin{tabular}[c]{@{}c@{}}303   GOPS/W (8b) \\      570 GOPS/W (4b)\\      1152 GOPS/W (2b)\end{tabular} & \begin{tabular}[c]{@{}c@{}}191 GOPS/W (8b)   \\      396 GOPS/W (4b)\\      835 GOPS/W (2b)\end{tabular} \\ \hline
\textit{\begin{tabular}[c]{@{}c@{}}Best   FP32\\ Performance\end{tabular}}     & --                                                              & --                                                               & 2 GFLOPS                                                                    & --                                                                                                        & 1.03 GFLOPS                                                                                              \\ \hline
\textit{\begin{tabular}[c]{@{}c@{}}Best FP32\\ Efficiency\end{tabular}}        & --                                                              & --                                                               & 79   GFLOPS/W                                                               & --                                                                                                        & 12 GFLOPS/W                                                                                              \\ \hline
\textit{\begin{tabular}[c]{@{}c@{}}Best   FP16 \\ Performance\end{tabular}}    & --                                                              & --                                                               & 3.3 GFLOPS                                                                  & --                                                                                                        & 18.2 GFLOPS                                                                                              \\ \hline
\textit{\begin{tabular}[c]{@{}c@{}}Best FP16 \\ Efficiency\end{tabular}}       & --                                                              & --                                                               & \begin{tabular}[c]{@{}c@{}}129   GFLOPS/W\\      @ 1.27 GFLOPS\end{tabular} & --                                                                                                        & \begin{tabular}[c]{@{}c@{}}300   GFLOPS/W\\      @ 2.6 GFLOPS\end{tabular}                              
\end{tabular}
}
\end{table*}

Tab.~\ref{tab:soa} compares \textsc{Darkside} with a wide range of programmable embedded computing platforms that exploit either parallelism or heterogeneity to address the computing requirements of emerging TinyML applications.

Compared to a traditional low-power programmable IoT system such as~\cite{bol2021sleeprunner},  representative of a wide range of low-cost microcontrollers embedding CortexM0, \textsc{Darkside} delivers several orders of magnitude better integer (8-bit) peak performance and also 1.9$\times$ better energy efficiency, despite \textit{SleepRunner}~\cite{bol2021sleeprunner} is implemented in a more scaled technology node (28nm FD-SOI). Contrarily to \textsc{Darkside}'s cluster, the implementation strategy of SleepRunner is highly optimized to operate at very low voltage (i.e. down to 0.4V). Its architecture features a simple memory hierarchy and interconnects scheme, which consumes very low power but poses severe limitations during the execution of complex near-sensor data analytic applications, which are efficiently sustained on \textsc{Darkside}.

With respect to hardware-accelerated IoT end-nodes such as \textit{SamurAI}~\cite{miro2020samurai}, implemented in 28nm FD-SOI technology, our SoC achieves similar energy efficiency on DNN workloads (only 1.2$\times$ less efficient despite the less scaled technology node used to implement Darkside, 65nm) but with a significant gain of 10$\times$ in terms of peak performance. This gain is primarily due to the custom extensions of  RV-NN cores and the parallel computing cluster over the sequential solution presented in~\cite{miro2020samurai}.

Finally, we compare \textsc{Darkside} with two SoCs that exploit a similar architectural template: Dustin~\cite{ottavi2022dustin} and Vega~\cite{VEGA} implement a multi-core RISC-V compute cluster in 65nm and 22nm, respectively. 
Compared to Vega~\cite{VEGA}, Darkside delivers better performance on 8-bit integer workloads thanks to the M\&L instruction. Contrarily to Vega, \textsc{Darkside} can support also mixed and lower-precision (than 8-bit) integer workloads thanks to the enhanced mixed-precision ISA, enabling the computation of emerging DNN models that employ asymmetric quantization schemes~\cite{rusci2020leveraging}. On 32-bit \ac{FP} workloads, Vega surpasses our solution in performance and energy efficiency due to the higher frequency operating mode and the much more scaled technology node. However, despite the previously mentioned advantages of Vega, the TPE of \textsc{Darkside} ensures 2.32$\times$ better energy efficiency on FP16 workloads, with a considerable performance gain of up to 5.6$\times$.

Compared to Dustin, featuring a cluster with 16 processors with mixed-precision extensions implemented in the same technology node, the proposed cluster shows slightly less energy efficiency due to the power reduction achieved by Dustin, thanks to the Vector Lockstep Execution Mode (VLEM)\footnote{The VLEM is not compatible with the cores unsed in Darkside (RVNN) and the two optimizations could be eventually combined to improve computing energy efficiency.}. However, \textsc{Darkside} still achieves 1.13$\times$ better performance with half of the cores, thanks to the M\&L extension. 

\section{Conclusion}
We presented \textsc{Darkside}, a low-power heterogeneous compute cluster for \textit{TinyML} DNN inference and on-chip training. The cluster features 8 RISC-V cores, enhanced with 2-bit to 32-bit mixed-precision integer SIMD instructions and fused mac-load operations. It also features specialized accelerators to boost the performance of integer depth-wise convolutions, reduce the latency of data marshalling operations, and enhance the performance and efficiency of FP16 kernels. The proposed SoC, implemented in TSMC 65nm technology, can achieve up to 65 GOPS peak performance on ML workloads, with 835 GOPS/W of energy efficiency. On FP16 kernels offloaded to the TPU, the SoC achieves 18.2 GFLOPS with 300 GFLOPS/W, surpassing the efficiency and performance of state-of-the-art SoCs implemented in much more scaled and expensive technology nodes.

\bibliographystyle{IEEEtran}
\bibliography{bibliography.bib}

\begin{thebibliography}{10}
\providecommand{\url}[1]{#1}
\csname url@samestyle\endcsname
\providecommand{\newblock}{\relax}
\providecommand{\bibinfo}[2]{#2}
\providecommand{\BIBentrySTDinterwordspacing}{\spaceskip=0pt\relax}
\providecommand{\BIBentryALTinterwordstretchfactor}{4}
\providecommand{\BIBentryALTinterwordspacing}{\spaceskip=\fontdimen2\font plus
\BIBentryALTinterwordstretchfactor\fontdimen3\font minus
  \fontdimen4\font\relax}
\providecommand{\BIBforeignlanguage}[2]{{%
\expandafter\ifx\csname l@#1\endcsname\relax
\typeout{** WARNING: IEEEtran.bst: No hyphenation pattern has been}%
\typeout{** loaded for the language `#1'. Using the pattern for}%
\typeout{** the default language instead.}%
\else
\language=\csname l@#1\endcsname
\fi
#2}}
\providecommand{\BIBdecl}{\relax}
\BIBdecl

\bibitem{sandler2018mobilenetv2}
M.~Sandler, A.~Howard, M.~Zhu, A.~Zhmoginov, and L.-C. Chen, ``{{Mobilenetv2:
  Inverted residuals and linear bottlenecks}},'' in \emph{Proceedings of the
  IEEE conference on computer vision and pattern recognition}, 2018, pp.
  4510--4520.

\bibitem{cai2018proxylessnas}
H.~Cai, L.~Zhu, and S.~Han, ``{{Proxylessnas: Direct neural architecture search
  on target task and hardware}},'' \emph{arXiv preprint arXiv:1812.00332},
  2018.

\bibitem{incze2018bird}
A.~Incze, H.-B. Jancs{\'o}, Z.~Szil{\'a}gyi, A.~Farkas, and C.~Sulyok, ``{{Bird
  sound recognition using a convolutional neural network}},'' in \emph{2018
  IEEE 16th International Symposium on Intelligent Systems and Informatics
  (SISY)}.\hskip 1em plus 0.5em minus 0.4em\relax IEEE, 2018, pp.
  000\,295--000\,300.

\bibitem{zhang2019lightweight}
B.~Zhang, Y.~Zhang, and S.~Wang, ``{{A lightweight and discriminative model for
  remote sensing scene classification with multidilation pooling module}},''
  \emph{IEEE Journal of Selected Topics in Applied Earth Observations and
  Remote Sensing}, vol.~12, no.~8, pp. 2636--2653, 2019.

\bibitem{ravaglia2021tinyml}
L.~Ravaglia \emph{et~al.}, ``{{A TinyML Platform for On-Device Continual
  Learning With Quantized Latent Replays}},'' \emph{IEEE Journal on Emerging
  and Selected Topics in Circuits and Systems}, vol.~11, no.~4, pp. 789--802,
  2021.

\bibitem{ren2021tinyol}
H.~Ren, D.~Anicic, and T.~A. Runkler, ``{{Tinyol: Tinyml with online-learning
  on microcontrollers}},'' in \emph{2021 International Joint Conference on
  Neural Networks (IJCNN)}.\hskip 1em plus 0.5em minus 0.4em\relax IEEE, 2021,
  pp. 1--8.

\bibitem{unknown}
C.~Banbury, V.~Janapa~Reddi, P.~Torelli, J.~Holleman, N.~Jeffries, C.~Kiraly,
  P.~Montino, D.~Kanter, S.~Ahmed, D.~Pau, U.~Thakker, A.~Torrini, P.~Warden,
  J.~Cordaro, G.~Di~Guglielmo, J.~Duarte, S.~Gibellini, V.~Parekh, H.~Tran, and
  X.~Xuesong, ``{{MLPerf Tiny Benchmark}},'' 06 2021.

\bibitem{jacob2018quantization}
B.~Jacob, S.~Kligys, B.~Chen, M.~Zhu, M.~Tang, A.~Howard, H.~Adam, and
  D.~Kalenichenko, ``{{Quantization and training of neural networks for
  efficient integer-arithmetic-only inference}},'' in \emph{Proceedings of the
  IEEE conference on computer vision and pattern recognition}, 2018, pp.
  2704--2713.

\bibitem{hubara2017quantized}
I.~Hubara, M.~Courbariaux, D.~Soudry, R.~El-Yaniv, and Y.~Bengio, ``{{Quantized
  neural networks: Training neural networks with low precision weights and
  activations}},'' \emph{The Journal of Machine Learning Research}, vol.~18,
  no.~1, pp. 6869--6898, 2017.

\bibitem{rusci2020leveraging}
M.~Rusci, M.~Fariselli, A.~Capotondi, and L.~Benini, ``{{Leveraging automated
  mixed-low-precision quantization for tiny edge microcontrollers}},'' in
  \emph{IoT Streams for Data-Driven Predictive Maintenance and IoT, Edge, and
  Mobile for Embedded Machine Learning}.\hskip 1em plus 0.5em minus 0.4em\relax
  Springer, 2020, pp. 296--308.

\bibitem{schaefer2022edge}
C.~J. Schaefer, S.~Joshi, S.~Li, and R.~Blazquez, ``{{Edge Inference with Fully
  Differentiable Quantized Mixed Precision Neural Networks}},'' \emph{arXiv
  preprint arXiv:2206.07741}, 2022.

\bibitem{chen2019eyeriss}
Y.-H. Chen, T.-J. Yang, J.~Emer, and V.~Sze, ``{{Eyeriss v2: A flexible
  accelerator for emerging deep neural networks on mobile devices}},''
  \emph{IEEE Journal on Emerging and Selected Topics in Circuits and Systems},
  vol.~9, no.~2, pp. 292--308, 2019.

\bibitem{moons201714}
B.~Moons, R.~Uytterhoeven, W.~Dehaene, and M.~Verhelst, ``{{14.5 envision: A
  0.26-to-10tops/w subword-parallel dynamic-voltage-accuracy-frequency-scalable
  convolutional neural network processor in 28nm fdsoi}},'' in \emph{2017 IEEE
  International Solid-State Circuits Conference (ISSCC)}.\hskip 1em plus 0.5em
  minus 0.4em\relax IEEE, 2017, pp. 246--247.

\bibitem{lee2018unpu}
J.~Lee, C.~Kim, S.~Kang, D.~Shin, S.~Kim, and H.-J. Yoo, ``{{UNPU: A 50.6
  TOPS/W unified deep neural network accelerator with 1b-to-16b fully-variable
  weight bit-precision}},'' in \emph{2018 IEEE International Solid-State
  Circuits Conference-(ISSCC)}.\hskip 1em plus 0.5em minus 0.4em\relax IEEE,
  2018, pp. 218--220.

\bibitem{desoli201714}
G.~Desoli, N.~Chawla, T.~Boesch, S.-p. Singh, E.~Guidetti, F.~De~Ambroggi,
  T.~Majo, P.~Zambotti, M.~Ayodhyawasi, H.~Singh \emph{et~al.}, ``{{14.1 A 2.9
  TOPS/W deep convolutional neural network SoC in FD-SOI 28nm for intelligent
  embedded systems}},'' in \emph{2017 IEEE International Solid-State Circuits
  Conference (ISSCC)}.\hskip 1em plus 0.5em minus 0.4em\relax IEEE, 2017, pp.
  238--239.

\bibitem{khaddam2022hermes}
R.~Khaddam-Aljameh, M.~Stanisavljevic, J.~F. Mas, G.~Karunaratne,
  M.~Br{\"a}ndli, F.~Liu, A.~Singh, S.~M. M{\"u}ller, U.~Egger, A.~Petropoulos
  \emph{et~al.}, ``{{HERMES-core—a 1.59-TOPS/mm 2 PCM on 14-nm CMOS in-memory
  compute core using 300-ps/LSB linearized CCO-based ADCs}},'' \emph{IEEE
  Journal of Solid-State Circuits}, vol.~57, no.~4, pp. 1027--1038, 2022.

\bibitem{ueyoshi2022diana}
K.~Ueyoshi, I.~A. Papistas, P.~Houshmand, G.~M. Sarda, V.~Jain, M.~Shi,
  Q.~Zheng, S.~Giraldo, P.~Vrancx, J.~Doevenspeck \emph{et~al.}, ``{{DIANA: An
  End-to-End Energy-Efficient Digital and ANAlog Hybrid Neural Network SoC}},''
  in \emph{2022 IEEE International Solid-State Circuits Conference (ISSCC)},
  vol.~65.\hskip 1em plus 0.5em minus 0.4em\relax IEEE, 2022, pp. 1--3.

\bibitem{garofalo2021xpulpnn}
A.~Garofalo \emph{et~al.}, ``{{XpulpNN: Enabling Energy Efficient and Flexible
  Inference of Quantized Neural Networks on RISC-V Based IoT End Nodes}},''
  \emph{IEEE Transactions on Emerging Topics in Computing}, vol.~9, no.~3, pp.
  1489--1505, 2021.

\bibitem{ottavi2020mixed}
G.~Ottavi \emph{et~al.}, ``{{A mixed-precision RISC-V processor for
  extreme-edge DNN inference}},'' in \emph{2020 IEEE Computer Society Annual
  Symposium on VLSI (ISVLSI)}.\hskip 1em plus 0.5em minus 0.4em\relax IEEE,
  2020, pp. 512--517.

\bibitem{garofalo2022heterogeneous}
A.~Garofalo \emph{et~al.}, ``{{A Heterogeneous In-Memory Computing Cluster For
  Flexible End-to-End Inference of Real-World Deep Neural Networks}},''
  \emph{arXiv preprint arXiv:2201.01089}, 2022.

\bibitem{rodriguez2018lower}
A.~Rodriguez \emph{et~al.}, ``{{Lower numerical precision deep learning
  inference and training}},'' \emph{Intel White Paper}, vol.~3, pp. 1--19,
  2018.

\bibitem{sun2019hybrid}
X.~Sun \emph{et~al.}, ``{{Hybrid 8-bit floating point (HFP8) training and
  inference for deep neural networks}},'' \emph{Advances in neural information
  processing systems}, vol.~32, 2019.

\bibitem{VEGA}
D.~Rossi \emph{et~al.}, ``{{Vega: A Ten-Core SoC for IoT Endnodes With DNN
  Acceleration and Cognitive Wake-Up From MRAM-Based State-Retentive Sleep
  Mode}},'' \emph{IEEE Journal of Solid-State Circuits}, vol.~57, no.~1, pp.
  127--139, 2021.

\bibitem{starwars}
G.~Lucas and G.~Kurtz, ``{{Star Wars Episode IV, A New Hope}},'' 1977.

\bibitem{burrello2021dory}
A.~Burrello, A.~Garofalo, N.~Bruschi, G.~Tagliavini, D.~Rossi, and F.~Conti,
  ``{{Dory: Automatic end-to-end deployment of real-world dnns on low-cost iot
  mcus}},'' \emph{IEEE Transactions on Computers}, vol.~70, no.~8, pp.
  1253--1268, 2021.

\bibitem{gautschi2017near}
M.~Gautschi, P.~D. Schiavone, A.~Traber, I.~Loi, A.~Pullini, D.~Rossi,
  E.~Flamand, F.~K. G{\"u}rkaynak, and L.~Benini, ``{{Near-threshold RISC-V
  core with DSP extensions for scalable IoT endpoint devices}},'' \emph{IEEE
  Transactions on Very Large Scale Integration (VLSI) Systems}, vol.~25,
  no.~10, pp. 2700--2713, 2017.

\bibitem{garofalo2020pulp}
A.~Garofalo, M.~Rusci, F.~Conti, D.~Rossi, and L.~Benini, ``Pulp-nn:
  accelerating quantized neural networks on parallel ultra-low-power risc-v
  processors,'' \emph{Philosophical Transactions of the Royal Society A}, vol.
  378, no. 2164, p. 20190155, 2020.

\bibitem{rakka2022mixed}
M.~Rakka, M.~E. Fouda, P.~Khargonekar, and F.~Kurdahi, ``Mixed-precision neural
  networks: A survey,'' \emph{arXiv preprint arXiv:2208.06064}, 2022.

\bibitem{redmuledate2022}
Y.~Tortorella, L.~Bertaccini, D.~Rossi, L.~Benini, and F.~Conti, ``{{RedMulE: A
  Compact FP16 Matrix-Multiplication Accelerator for Adaptive Deep Learning on
  RISC-V-Based Ultra-Low-Power SoCs}},'' 2022.

\bibitem{nvidia_mixp}
\BIBentryALTinterwordspacing
``{{Training With Mixed Precision - NVIDA Deep Learning Performance
  Documentation}},'' 2021. [Online]. Available:
  \url{https://docs.nvidia.com/deeplearning/performance/mixed-precision-training/index.html#tensorop}
\BIBentrySTDinterwordspacing

\bibitem{tagliavini2018transprecision}
G.~Tagliavini, S.~Mach, D.~Rossi, A.~Marongiu, and L.~Benini, ``{{A
  transprecision floating-point platform for ultra-low power computing}},'' in
  \emph{2018 Design, Automation \& Test in Europe Conference \& Exhibition
  (DATE)}.\hskip 1em plus 0.5em minus 0.4em\relax IEEE, 2018, pp. 1051--1056.

\bibitem{meloni2018neuraghe}
P.~Meloni, A.~Capotondi, G.~Deriu, M.~Brian, F.~Conti, D.~Rossi, L.~Raffo, and
  L.~Benini, ``Neuraghe: Exploiting cpu-fpga synergies for efficient and
  flexible cnn inference acceleration on zynq socs,'' \emph{ACM Transactions on
  Reconfigurable Technology and Systems (TRETS)}, vol.~11, no.~3, pp. 1--24,
  2018.

\bibitem{bruschi2021gvsoc}
N.~Bruschi, G.~Haugou, G.~Tagliavini, F.~Conti, L.~Benini, and D.~Rossi,
  ``{{GVSoC: A Highly Configurable, Fast and Accurate Full-Platform Simulator
  for RISC-V based IoT Processors}},'' in \emph{2021 IEEE 39th International
  Conference on Computer Design (ICCD)}.\hskip 1em plus 0.5em minus 0.4em\relax
  IEEE, 2021, pp. 409--416.

\bibitem{nando2022samos}
D.~Nadalini, M.~Rusci, G.~Tagliavini, L.~Ravaglia, L.~Benini, and F.~Conti,
  ``{{PULP-TrainLib: Enabling On-Device Training for RISC-V Multi-Core MCUs
  through Performance-Driven Autotuning}},'' in \emph{to appear at 2022
  Springer 22nd International Conferenceon Embedded Computer Systems:
  Architectures, Modeling and Simulation, SAMOS XXII}.\hskip 1em plus 0.5em
  minus 0.4em\relax Springer, 2022, p.~1.

\bibitem{bol2021sleeprunner}
D.~Bol \emph{et~al.}, ``{{SleepRunner: A 28-nm FDSOI ULP Cortex-M0 MCU With ULL
  SRAM and UFBR PVT Compensation for 2.6--3.6-$\mu$W/DMIPS 40--80-MHz Active
  Mode and 131-nW/kB Fully Retentive Deep-Sleep Mode}},'' \emph{IEEE Journal of
  Solid-State Circuits}, vol.~56, no.~7, pp. 2256--2269, 2021.

\bibitem{miro2020samurai}
Miro-Panades \emph{et~al.}, ``{{SamurAI: a 1.7MOPS-36GOPS Adaptive Versatile
  IoT Node with 15,000x Peak-to-Idle Power Reduction, 207ns Wake-up Time and
  1.3TOPS/W ML Efficiency}},'' in \emph{2020 IEEE Symposium on VLSI
  Circuits}.\hskip 1em plus 0.5em minus 0.4em\relax IEEE, 2020, pp. 1--2.

\bibitem{ottavi2022dustin}
G.~Ottavi \emph{et~al.}, ``{{Dustin: A 16-Cores Parallel Ultra-Low-Power
  Cluster with 2b-to-32b Fully Flexible Bit-Precision and Vector Lockstep
  Execution Mode}},'' \emph{arXiv preprint arXiv:2201.08656}, 2022.

\end{thebibliography}

\begin{IEEEbiography}[{\includegraphics[width=1in,height=1.25in,clip,keepaspectratio]{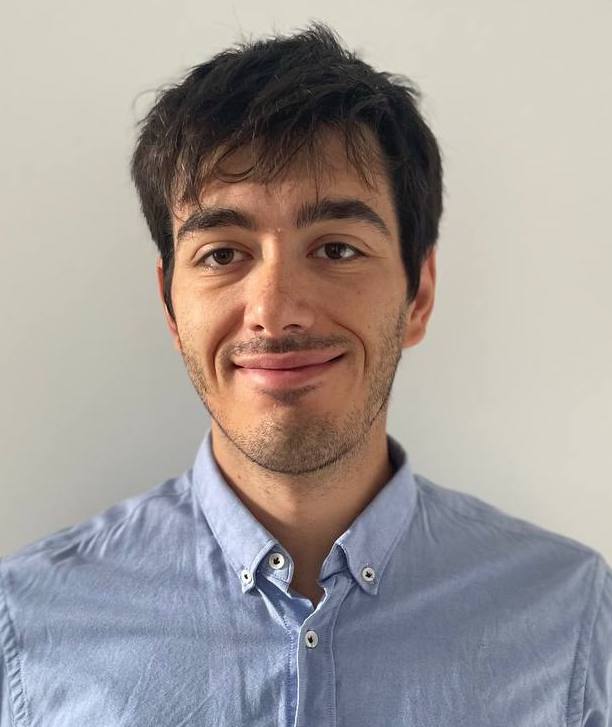}}]{Angelo Garofalo} received the B.Sc and M.Sc. degree in electronic engineering from the University of Bologna, Italy, in 2016 and 2018 respectively. He is currently working toward his Ph.D. degree at DEI, University of Bologna, Italy. His main research topic is on Flexible Computing Systems for AI Acceleration at the Extreme Edge of the IoT. His research interests include Quantized Neural Networks, Hardware efficient Machine Learning, in-memory computing heterogeneous architectures and fully-programmable embedded architectures.
\end{IEEEbiography}
\begin{IEEEbiography}[{\includegraphics[width=1in,height=1.25in,clip,keepaspectratio]{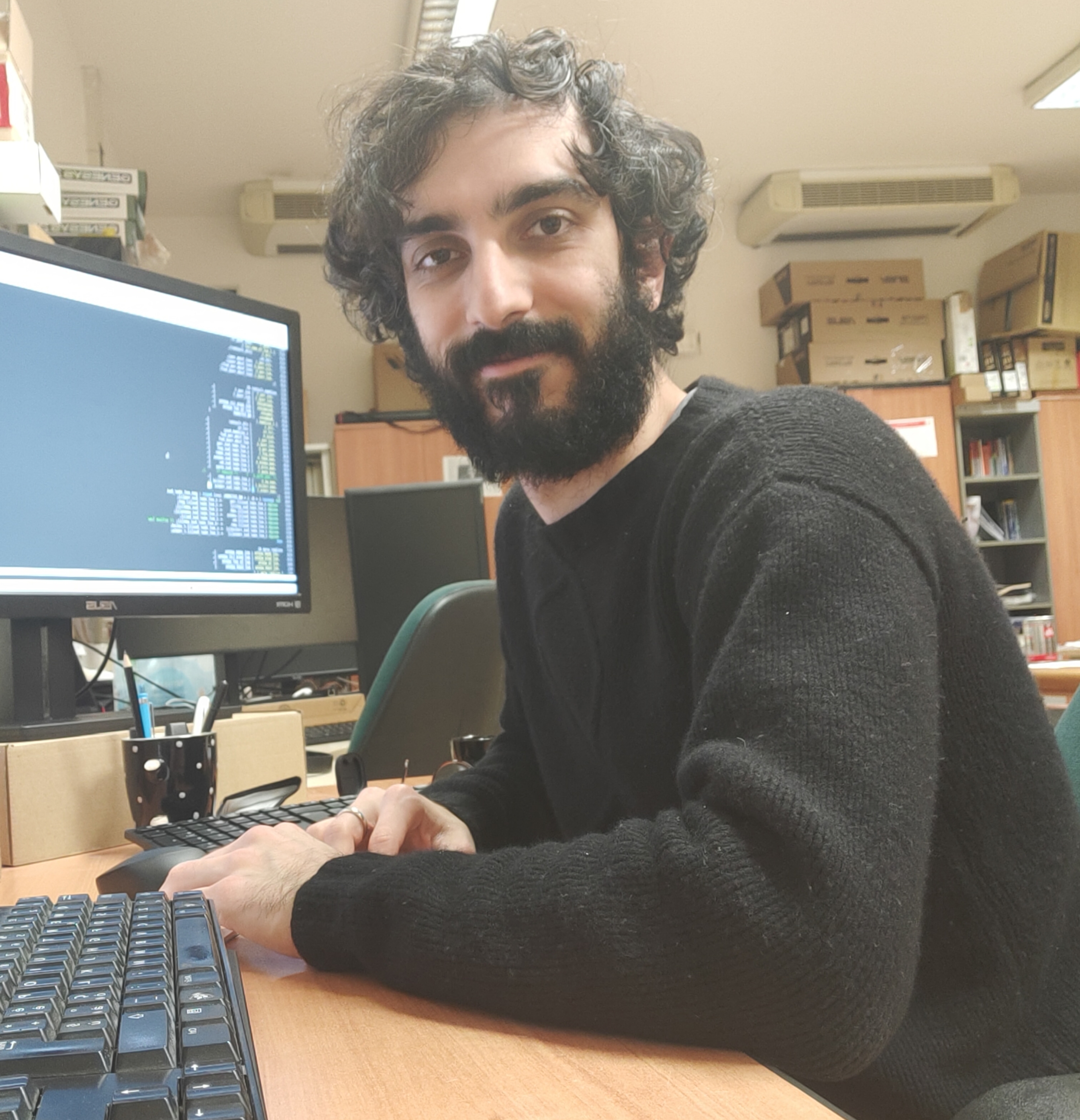}}]{Yvan Tortorella} received his Master's Degree in Electronic Engineering in October 2021 from the University of Bologna. He is currently pursuing a Ph. D. in Digital Systems Design in the group of Professor Luca Benini at the Department of Electrical and Information Engineering (DEI) of the University of Bologna. His research interests include the design of PULP (Parallel Ultra-Low Power)-based hardware accelerators for ultra-low power Machine Learning and the design of RISC-V-based computer architectures for satellite applications.
\end{IEEEbiography}

\begin{IEEEbiography}[{\includegraphics[width=1in,height=1.25in,clip,keepaspectratio]{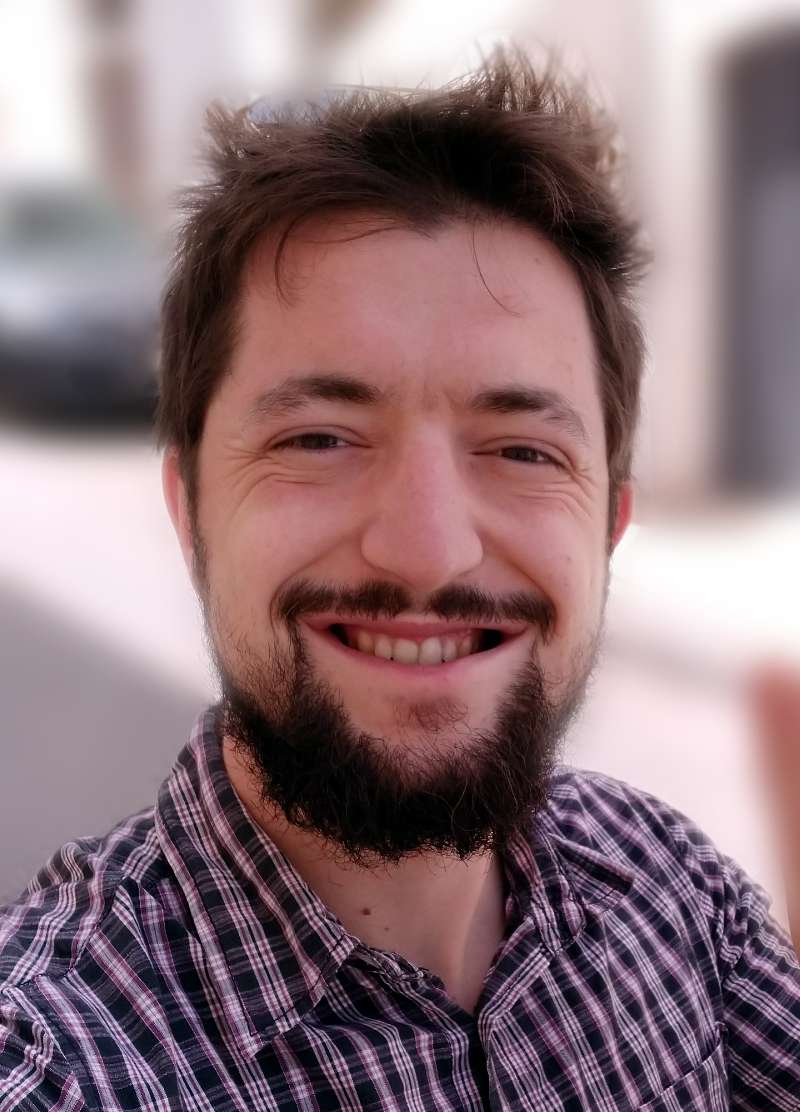}}] {Matteo Perotti} received his M.Sc. degree in Electronic Engineering from the Polytechnic University of Turin, Italy, in 2019. He is currently pursuing a Ph.D. degree at the Integrated Systems Laboratory of ETH Zurich, Switzerland. His research interests include highly efficient compute architectures and computation with high dynamic-range data types.
\end{IEEEbiography}

\begin{IEEEbiography}[{\includegraphics[width=1in,height=1.25in,clip,keepaspectratio]{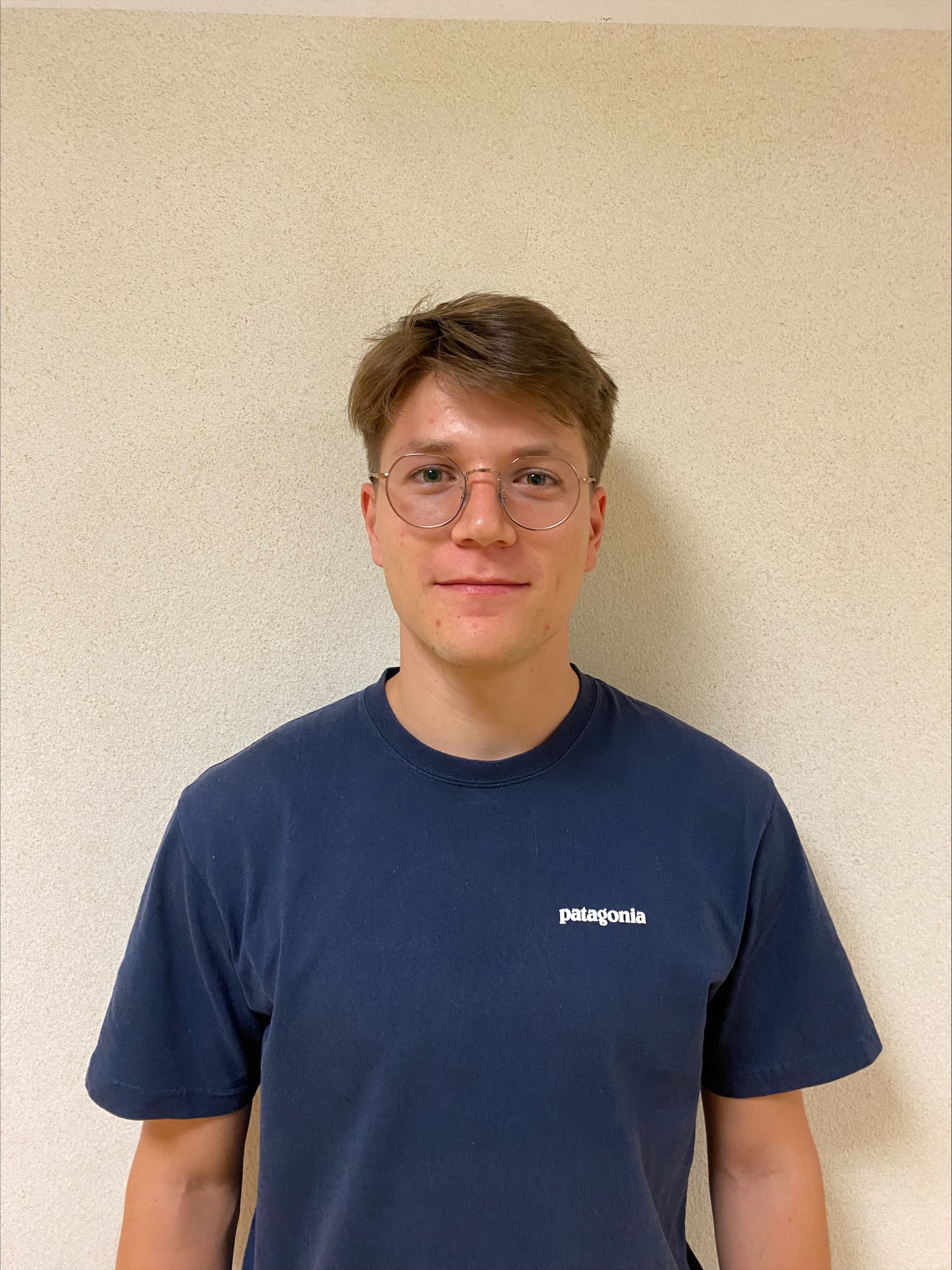}}] {Luca Valente} received the MSc degree in electronic engineering from the Politecnico of Turin in 2020. He is currently a PhD student at the University of Bologna within the Department of Electrical, Electronic and Information
Technologies Engineering (DEI). His main research interests are hardware-software co-design of multi-processors heterogenous systems on chip, parallel programming and FPGA prototyping.
\end{IEEEbiography}

\begin{IEEEbiography}[{\includegraphics[width=1in,height=1.25in,clip,keepaspectratio]{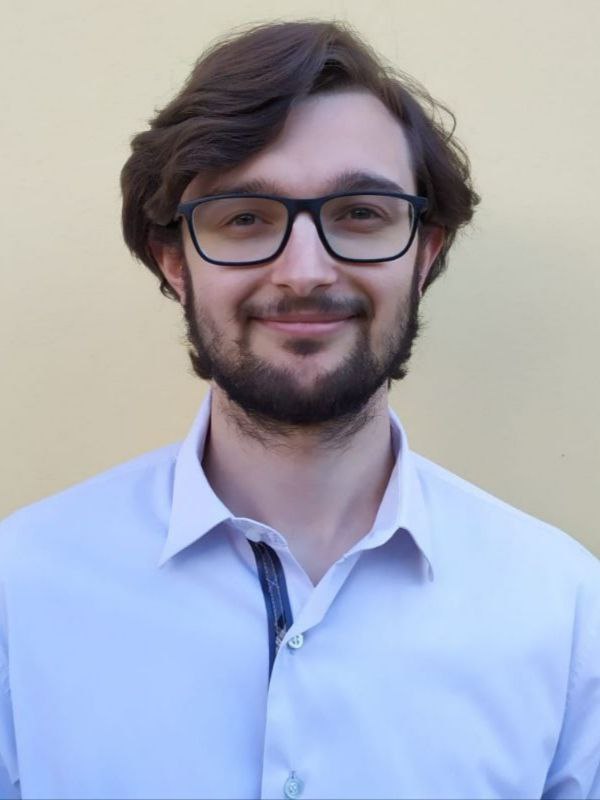}}]
{Alessandro Nadalini} received the B.Sc. and M.Sc. degrees in electronic engineering from the University of Bologna, Italy, in 2018 and 2021, respectively. He currently holds a research grant from the University of Bologna, Bologna, Italy. His research regards lightweight extensions to the RISC-V ISA to boost the efficiency of heavily Quantized Neural Networks inference on microcontroller-class cores.
\end{IEEEbiography}

\begin{IEEEbiography}[{\includegraphics[width=1in,height=1.25in,clip,keepaspectratio]{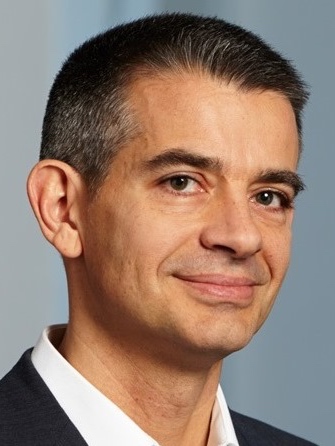}}]{Luca Benini} holds the chair of digital Circuits and systems at ETHZ and is Full Professor at the Università di Bologna. He received a PhD from Stanford University. Dr. Benini's research interests are in energy-efficient parallel computing systems, smart sensing micro-systems and machine learning hardware. He has published more than 1000 peer-reviewed papers and five books. He is a Fellow of the ACM and a member of the Academia Europaea.
\end{IEEEbiography}
\begin{IEEEbiography}[{\includegraphics[width=1in,height=1.25in,clip,keepaspectratio]{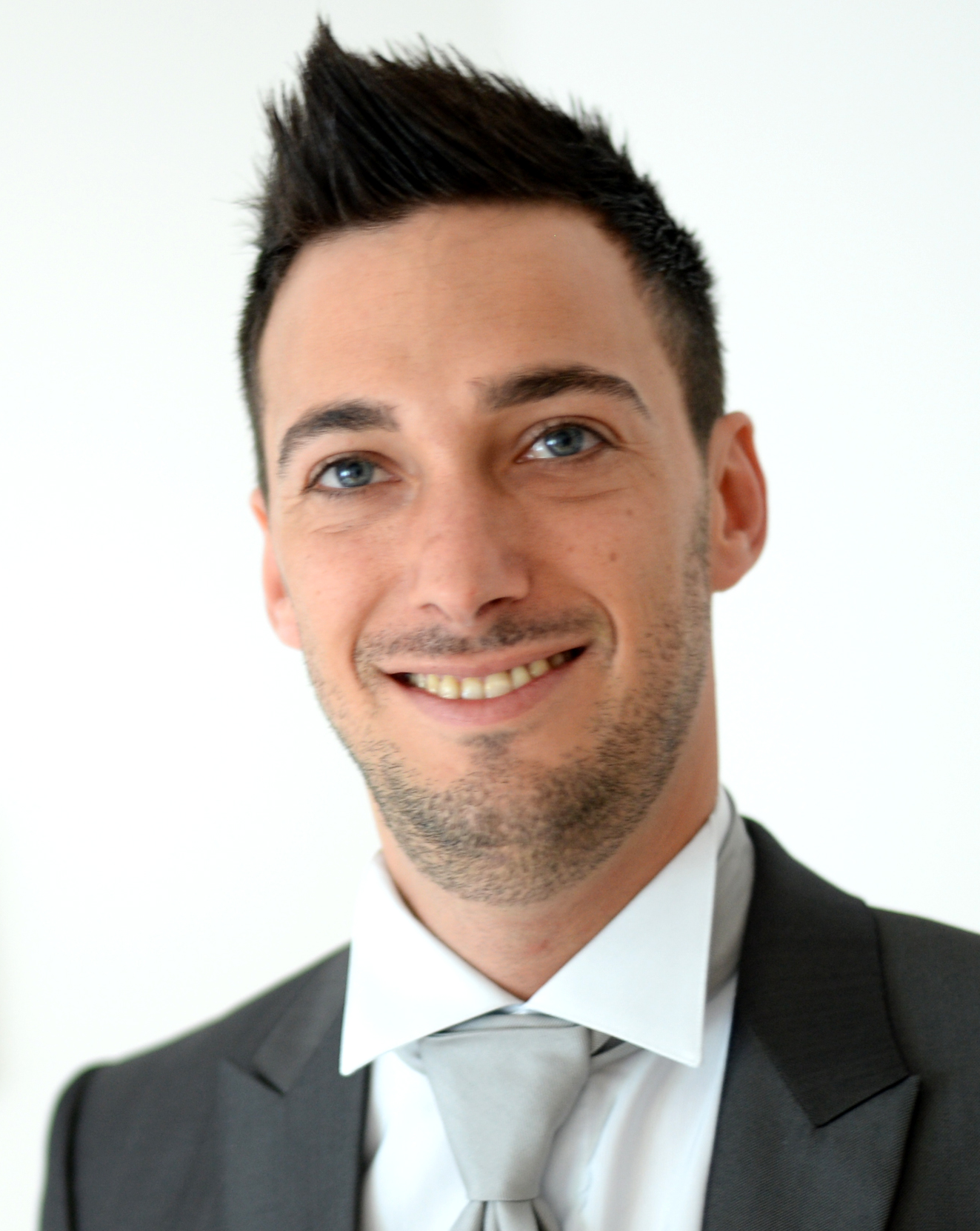}}]{Davide Rossi} received the Ph.D. degree from the University of Bologna, Bologna, Italy, in 2012. He has been a Post-Doctoral Researcher with the Department of Electrical, Electronic and Information Engineering “Guglielmo Marconi,” University of Bologna, since 2015, where he is currently an Assistant Professor. His research interests focus on energy-efficient digital architectures. In this field, he has published more than 100 papers in international peer-reviewed conferences and journals. He is recipient of Donald O. Pederson Best Paper Award 2018, 2020 IEEE TCAS Darlington Best Paper Award, 2020 IEEE TVLSI Prize Paper Award.
\end{IEEEbiography}
\begin{IEEEbiography}[{\includegraphics[width=1in,height=1.25in,clip,keepaspectratio]{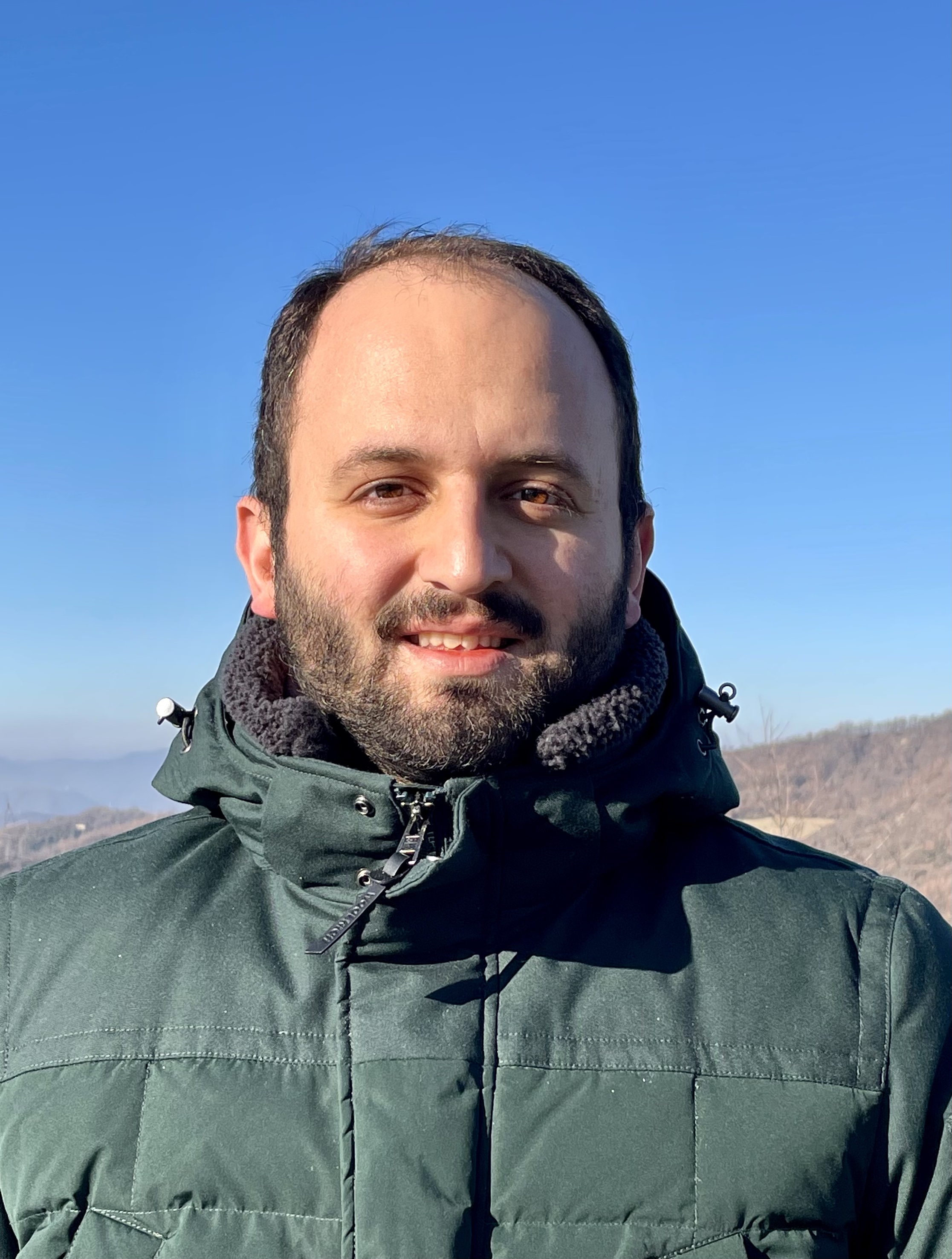}}]{Francesco Conti} received the Ph.D. degree in electronic engineering from the University of Bologna, Italy, in 2016. He is currently an Assistant Professor in the DEI Department of the University of Bologna. From 2016 to 2020, he held a research grant in the DEI department of University of Bologna and a position as postdoctoral researcher at the Integrated Systems Laboratory of ETH Zurich in the Digital Systems group. His research focuses on the development of deep learning based intelligence on top of ultra-low power, ultra-energy efficient programmable Systems-on-Chip. His research work has resulted in more than 40 publications in international conferences and journals and has been awarded several times, including the 2020 IEEE TCAS-I Darlington Best Paper Award.
\end{IEEEbiography}

\end{document}